\documentclass[11pt]{article}
 
\usepackage[]{acl}
 
\usepackage{times}
\usepackage{latexsym}
\usepackage{booktabs,multirow,longtable}
\usepackage[T1]{fontenc}
 
\usepackage[utf8]{inputenc}

\usepackage{microtype}
\usepackage{graphicx}
\usepackage{subcaption}
\usepackage{inconsolata}
 
\usepackage{graphicx}
 
\usepackage{multirow}
\usepackage{colortbl}
\usepackage{xcolor}
\usepackage{bm}
\usepackage{enumitem}
\usepackage{mathtools}
\usepackage{booktabs}
\usepackage{comment}
\usepackage{tcolorbox}
\tcbuselibrary{breakable, skins}
\usepackage{enumitem}
\usepackage{float}
\usepackage{algorithm}
\usepackage{algpseudocode}

\usepackage{microtype}
\usepackage{microtype}
\usepackage{hyperref}
\usepackage{url}
\usepackage{booktabs}
\usepackage{tabularx}
\usepackage{multirow}
\usepackage{tcolorbox}
\usepackage{enumitem}
\usepackage{xcolor}
\usepackage{amssymb}
\usepackage{subcaption}
\usepackage{xcolor}
\usepackage{colortbl}

\newcommand{\pos}[1]{#1}
\newcommand{\neu}[1]{\textcolor{gray}{#1}}

\usepackage{pgfplots}
\pgfplotsset{compat=1.18}
\usepackage{pgfplotstable}
\usepackage{graphicx}
\usepackage{wrapfig}
\usepackage{lineno}
\usepackage{amsmath}
\usepackage{booktabs}
\usepackage[table]{xcolor}
\usepackage{arydshln}
\usepackage{booktabs}
\usepackage{multirow}
\usepackage{subcaption}
\usepackage{placeins}
\usepackage{xcolor}
\usepackage[export]{adjustbox}
\usepackage{colortbl}
\usepackage{rotating}   
\usepackage{array}
\usepackage{makecell}
\definecolor{darkblue}{rgb}{0, 0, 0.5}
\hypersetup{colorlinks=true, citecolor=darkblue, linkcolor=darkblue, urlcolor=darkblue}

\title{ RAG over Thinking Traces Can Improve Reasoning Tasks
}

\author{Negar Arabzadeh \qquad Wenjie Ma \qquad Sewon Min \qquad Matei Zaharia \\[1ex]
University of California, Berkeley \\[1ex]
}
\newcommand{\method}{\ensuremath{\mathcal{T}^3}}

\newcommand{\sewon}[1]{{\color{violet} [Sewon]: #1}}

\begin{document}






\let\inputSpace\mlInputs
\let\instance\mlInput
\let\outputSpace\mlOutputs

\newcommand{\mlOutput}{\hat{y}}
\let\output\mlOutput

\newcommand{\mlTarget}{y}
\let\target\mlTarget

\newcommand{\query}{q}
\newcommand{\querySpace}{\mathcal{Q}}

\newcommand{\rModel}{\mathcal{R}}

\newcommand{\llmModel}{\mathcal{G}}
\newcommand{\queryGenerator}{\phi_q}
\newcommand{\promptGenerator}{\phi_p}
\newcommand{\augmentedPrompt}{\overline{\instance}}

\newcommand{\doc}{d}
\newcommand{\corpus}{\mathcal{D}}

\newcommand{\retrievalResult}{z}
\newcommand{\retrievalResultSpace}{\mathcal{Z}}

\let\ranking\raRanking

\newcommand{\retrievalScores}{\mathbf{s}}
\newcommand{\retrievalScore}{s}

\newcommand{\utilityMetric}{\evalMetric_u}
\newcommand{\relevanceMetric}{\evalMetric_r}

\newcommand{\agent}{A}
\newcommand{\env}{E}

\newcommand{\trajectory}{\mathcal{T}}
\newcommand{\reasoning}{\mathbf{r}}
\newcommand{\action}{\mathbf{a}}
\newcommand{\observation}{\mathbf{o}}

\newcommand{\goal}{g}

\newcommand{\missing}[0]{{\color{red}[CITE]}}
\newcommand{\todo}[1]{\textcolor{red}{[TODO: #1]}}

\newcommand{\matei}[1]{\textcolor{green}{[matei: #1]}}
\newcommand{\wenjie}[1]{\textcolor{blue}{[wenjie: #1]}}
\renewcommand{\sewon}[1]{\textcolor{cyan}{[sewon: #1]}}
\newcommand{\negar}[1]{\textcolor{magenta}{[negar: #1]}}

\let\oldparagraph\paragraph
\renewcommand{\paragraph}[1]{\vspace{-.5em}
\oldparagraph{#1}
}

\makeatletter
\newcommand*\myfontsize{%
  \@setfontsize\myfontsize{7.7}{9}%
}
\makeatother

\newcommand{\mytextbox}[2]{\tikzmarknode[draw=#1,thick,inner sep=2pt]{test}{\myfontsize #2}}

\newcommand{\tagStruc}{\textcolor{orange}{\scriptsize Struct }}
\newcommand{\tagSem}{\textcolor{violet}{\scriptsize Semantic }}
\newcommand{\tagRefl}{\textcolor{cyan!70!black}{\scriptsize Reflect }}

\maketitle

\begin{abstract}
Retrieval-augmented generation (RAG) has proven effective for knowledge-intensive tasks, but is widely believed to offer limited benefit for reasoning-intensive problems such as math and code generation. We challenge this assumption by showing that the limitation lies not in RAG itself, but in the choice of corpus. Instead of retrieving documents, we propose retrieving thinking traces, i.e., intermediate thinking trajectories generated during problem solving attempts. We show that thinking traces are already a strong retrieval source, and further introduce \method, an offline method that transforms them into structured, retrieval-friendly representations, to improve usability.
Using these traces as a corpus, a simple retrieve-then-generate pipeline consistently improves reasoning performance across strong models and benchmarks such as AIME 2025--2026, LiveCodeBench, and GPQA-Diamond, outperforming both non-RAG baselines and retrieval over standard web corpora. For instance, on AIME 2025-2026, RAG with traces generated by Gemini-2-thinking achieves relative gains of +56.3\%, +8.6\%, and +7.6\% for Gemini-2.5-Flash, GPT-OSS-120B, and GPT-5, respectively, even though these are more recent models. 
Overall, our results suggest that thinking traces are an effective retrieval corpus for reasoning tasks, and transforming them into structured, compact, or diagnostic representations unlocks even stronger gains.

\end{abstract}

\section{Introduction}
\label{sec:introduction}

Retrieval-augmented generation (RAG) has become a standard way to improve large language models (LLMs) on knowledge-intensive tasks by retrieving external documents that provide factual grounding \citep{lewis2020retrieval,fan2024survey}. However, its value for reasoning-intensive tasks remains far less clear. Prior work suggests that standard retrieval over general-purpose corpora often provides limited or inconsistent gains for tasks such as mathematical reasoning, that these gains tend to appear mainly for weaker models \citep{lyu2025frustratinglysimpleretrievalimproves}, and that it can even hurt performance when the retrieved context is noisy or poorly aligned with the reasoning process \citep{li2025can,shi2023large,geng2024greatmemoryshallowreasoning,behnamghader2023can}. This has contributed to a growing belief that retrieval may be less helpful for reasoning than it is for factual question answering \citep{gao2023retrieval}.

In this work, we challenge the assumption that RAG is ineffective for reasoning \citep{liu2024much,lyu2025frustratinglysimpleretrievalimproves}, arguing that the limitation lies not in retrieval itself but in the choice of the retrieval corpus.
While prior RAG predominantly uses knowledge sources or generic web or textbook as retrieval corpora, they have shown to be better suited for factual recall than for reasoning tasks such as math. Instead, we posit that reasoning benefits from access to process-level signals such as how solutions are derived.
Motivated by this, we propose using \emph{thinking traces}—intermediate reasoning trajectories generated during problem-solving by state-of-the-art reasoning models—as a retrieval corpus for reasoning-focused RAG. 
We find that simply replacing standard web corpora with raw thinking traces already yields surprising gains for reasoning tasks.

\begin{figure*}[t]
\centering
\vspace{-1em}
 \includegraphics[width=\textwidth]{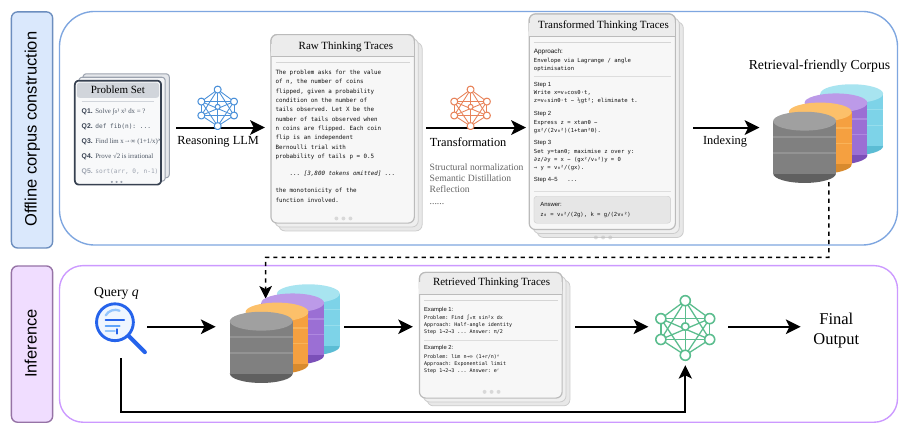}
\caption{%
{Overview of \method.} Offline, a large reasoning model (e.g., Gemini-2-thinking) solves a set of problems and produces raw thinking traces. A smaller model (e.g., Gemini-2-Flash-Lite) then rewrites them into structured representations, forming a retrieval-friendly corpus. At inference time, a previously unseen query, which is not part of initial problem set, is retrieved against this corpus, and the retrieved context is provided to a downstream LLM to generate the final answer. The inference model may differ from the trace-generation and transformation models.}
\vspace{-1em}
\label{fig:overview}
\end{figure*}



At the same time, na\"ively retrieving raw traces is suboptimal: full thinking traces from the state-of-the-art reasoning models are often lengthy, noisy and redundant, making them difficult for downstream models to use effectively. We therefore propose \method\ ($\textbf{T}$ransformation of $\textbf{T}$hinking $\textbf{T}$races), an offline method that transforms thinking traces into more structured, retrieval- and context-friendly forms. Rather than providing raw reasoning trajectories, \method\ distills them into concise scaffolds that provide a ``how-to'' for the reasoning process rather than mere factual grounding.
More broadly, we treat thinking traces as a reusable resource and rather than distilling them into model parameters or discarding them after inference, we transform and retrieve them for future problems.
In this sense, our setup is closer to learning from others' prior reasoning attempts and mistakes than to revising a model's own reasoning online. Because the trace corpus is built from a fully separate auxiliary problem set and generated by different models than those used at inference time, it also remains cleanly separated from the evaluation queries and reduces the risk of contamination.

Figure~\ref{fig:overview} illustrates the overview of the process. In an offline stage, a strong \emph{thinking model} (e.g., Gemini-2-thinking) generates thinking traces over a curated problem set, which a smaller \emph{transformation model} (e.g., Gemini-2-Flash-Lite) rewrites into retrieval-friendly forms using \method.
These transformed traces form the retrieval corpus.
At inference time, a standard retrieve-then-generate pipeline retrieves relevant trace segments and conditions a \emph{solver model} on them, replacing conventional web or knowledge corpora with transformed thinking traces.
The thinking, transformation, and solver models may be identical or distinct. Notably, our experiments show that even weaker thinking or transformation models  can significantly improve stronger solver models.


We run experiments across multiple frontier models, including GPT-OSS-120B \citep{openai2025gptoss120bgptoss20bmodel}, GPT-5 \citep{singh2025openaigpt5card}, and Gemini-2.5-Flash \citep{team2023gemini}, and across reasoning-intensive benchmarks spanning math(AIME 2025--2026), coding (LiveCodeBench) \citep{jain2024livecodebenchholisticcontaminationfree}, and scientific question answering (GPQA-Diamond) \citep{rein2023gpqagraduatelevelgoogleproofqa}. To test whether these gains extend beyond frontier-scale models, we further evaluate \method{} on several substantially smaller open models in Appendix~\ref{app:smaller_models}. Our contributions in this work are as follows:

\begin{enumerate}[leftmargin=*]
\item \textbf{We show that raw thinking traces are a uniquely effective retrieval source for reasoning-intensive tasks.} On AIME 2025--2026, raw-trace retrieval on Gemini-2-thinking improves Gemini-2.5-Flash from 53.3 to 80.0 (+50.1\%), GPT-OSS-120B from 78.3 to 85.0 (+8.6\%), and GPT-5 from 86.7 to 91.7 (+5.8\%). 
\newcommand{\yes}{\textcolor[HTML]{2e7d32}{$\checkmark$}}
\newcommand{\no}{\textcolor[HTML]{c62828}{$\times$}}

\begin{table*}[t]
\centering
\small
\setlength{\tabcolsep}{4pt}
\renewcommand{\arraystretch}{1}
\scalebox{0.8}{
\begin{tabular}{lcccc}
\toprule
\textbf{Method} 
& \makecell{\textbf{Retrieved object}} 
& \makecell{\textbf{Reusable artifact}} 
& \makecell{\textbf{Inference-time mechanism}} 
& \makecell{\textbf{Vanilla RAG}} \\
\midrule
RAT~\citep{wang2024ratretrievalaugmentedthoughts}
& Documents
& \no
& Iterative CoT revision
& \no \\
TRACE~\citep{fang-etal-2024-trace}
& Documents/chains
& \no
& Evidence-chain construction
& \no \\
BoT~\citep{yang2024buffer}
& Thought templates
& Meta-buffer~\yes
& Template instantiation
& \no \\
RoT~\citep{ahmed2025retrievalofthoughtefficientreasoningreusing}
& Thought graph nodes
& Thought graph~\yes
& Reward-guided traversal
& \no \\

Dynamic Cheatsheet~\citep{suzgun2026dynamic}
& Prior solutions/strategies
& Evolving cheatsheet~\yes
& Sequential memory curation
& \no \\
Reasoning Memory~\citep{wu2026proceduralknowledgescaleimproves}
& Subquestion-subroutine pairs
& Procedural datastore~\yes
& In-thought retrieval
& \no \\

\method{}
& Transformed thinking traces
& Trace-derived corpus~\yes
& Simple retrieval
& \yes \\
\bottomrule
\end{tabular}}
\caption{
Comparison with closely related reasoning-RAG systems.
}
\label{tab:related-work-compare}
\end{table*}

\item \textbf{We propose \method, an offline method for transforming thinking traces into more retrieval-friendly representations.} \method\ converts raw traces generated by strong reasoning models, including QwQ-32B and Gemini-2-thinking, into structured and more usable forms using a relatively light LLM (e.g., Gemini-2-Flash-Lite). In several cases where raw traces yield limited gains, transformed traces unlock clear improvements across tasks. For example, on GPQA Diamond, \method\ improves GPT-OSS-120B from 70.7 to 74.7 (+5.7\%), and on LiveCodeBench from 57.9 to 61.4 (+6.0\%).

\item \textbf{We show that RAG on \method{} can improve the cost--accuracy trade-off.} 
By shifting computation from expensive test-time decoding to cheaper input context, retrieval over thinking traces not only can improve answer quality, but in the best setting, it also makes inference cost up to $15\%$ cheaper per query (e.g., GPT-5).
\end{enumerate}

We release the code and data used in this paper to support future research at \url{https://github.com/Narabzad/t3}.

\section{Related Work}
\oldparagraph{Reasoning in LLMs.}

LLMs have recently shown strong performance on reasoning-intensive tasks \citep{wang2025surveylargelanguagemodels,rozière2024codellamaopenfoundation,auer2023sciqa}. Prior work improves reasoning through prompting and chain-of-thought reasoning \citep{wei2022chain,wang2022self}, distillation from stronger reasoning models \citep{ho2023largelanguagemodelsreasoning,magister2023teachingsmalllanguagemodels,shridhar2023distilling,muennighoff2025s1simpletesttimescaling}, and reinforcement learning with verifiable rewards \citep{guo2025deepseek,yu2025dapoopensourcellmreinforcement}. Recent work also studies how reasoning can interact with retrieval- or memory-like mechanisms \citep{wang2026survey,du2025memr3memoryretrievalreflective}. Our work is complementary: instead of internalizing reasoning through training or distillation, we study whether prior thinking traces can be stored externally and retrieved at inference time to guide reasoning.

\paragraph{RAG for Reasoning.} 
RAG is widely used for knowledge-intensive tasks by retrieving documents that provide factual grounding \citep{lewis2020retrieval,gao2023retrieval,siriwardhana2023improving}. Most work focuses on retrieving, selecting, and incorporating textual evidence from large corpora \citep{fan2024survey,singal2024evidence,huo2023retrieving}, with recent work also studying retrieval as an inference-time scaling mechanism where larger datastores can improve retrieval-based language models \citep{shao2024scalingretrievalbasedlanguagemodels}. While effective for factual question answering, its benefits for reasoning-intensive tasks remain less clear and unreliable.This motivates methods that couple retrieval with intermediate reasoning traces.

Several recent systems reuse ``thoughts'' or reasoning artifacts, but these usually refer to generated chains, templates, or graph nodes rather than the native long-form thinking traces produced by reasoning models. 
Retrieval-Augmented Thoughts (RAT) \citep{wang2024ratretrievalaugmentedthoughts} revises an initial chain of thought step by step using retrieved information, showing retrieval can help when tightly coupled with generation;
TRACE \citep{fang-etal-2024-trace} constructs knowledge-grounded reasoning chains from retrieved documents  using additional LLM calls before answer generation;  Buffer of Thoughts (BoT) \citep{yang2024buffer} retrieves high-level thought templates from a meta-buffer and instantiates them for the current task. Retrieval-of-Thought (RoT) \citep{ahmed2025retrievalofthoughtefficientreasoningreusing} organizes reusable reasoning steps into a thought graph and uses reward-guided traversal to assemble problem-specific templates at inference time.
Other work improves reasoning retrieval by training specialized retrievers \citep{shao2025reasonirtrainingretrieversreasoning}, scaling general-purpose datastores \citep{lyu2025frustratinglysimpleretrievalimproves}, or retrieving procedural subroutines within the reasoning process itself \citep{wu2026proceduralknowledgescaleimproves}. Closely related, Dynamic Cheatsheet \citep{suzgun2026dynamic} maintains a persistent, evolving memory of strategies curated online across a sequence of test-time queries, rather than built offline from existing thinking traces as in our work. Table~\ref{tab:related-work-compare} summarizes these differences.

Rather than proposing a new online reasoning controller, we ask whether reasoning gains can be obtained by changing the retrieval corpus while keeping the inference pipeline fixed. This preserves the simplicity, cost profile, and deployability of standard retrieve-then-generate RAG, requiring no additional inference-time model, controller, graph traversal, template instantiation, or iterative revision loop.
In contrast to the closest prior work summarized in Table~\ref{tab:related-work-compare}, our new component is offline corpus construction. To our knowledge, we are the first to study how existing thinking traces can be transformed into a reusable retrieval corpus for reasoning. We provide an expanded discussion of prior work in Appendix~\ref{app:related_work}.
\section{Methodology}

\label{sec:method}

\newcommand{\mycomment}[1]{\textcolor{blue!70!black}{\textit{#1}}}

\begin{figure*}[t]
\vspace{-3.5em}
\centering

\fcolorbox{black!20}{gray!3}{
\begin{minipage}{0.92\linewidth}

{\scriptsize\ttfamily
\setlength{\tabcolsep}{4pt}
\renewcommand{\arraystretch}{1.17}

\begin{tabular}{@{}p{\linewidth}@{}}
\toprule

\textbf{Problem (AIME 2026):} Call finite sets of integers $S$ and $T$ \emph{cousins} if they have the same number of elements, are disjoint, and their elements can be paired so that each pair differs by exactly $1$. Suppose that $S$ has exactly $4040$ cousins. Find the least number of elements $S$ can have. \\

\midrule
\rowcolor{gray!10} 
\textbf{Without retrieval} \hfill \textcolor{red}{0/8} \\
\mycomment{The model starts correctly --- it identifies that $S$ cannot contain three consecutive integers and partitions $S$ into isolated and paired blocks --- but never completes the argument. All 8 samples terminate mid-sentence before reaching a final answer.}\\[2pt]
\midrule
\rowcolor{gray!10} 
\textbf{Retrieved context from \method- \tagRefl} \\

\textbf{Example 1.} \textit{Set theory/Real Analysis} --- Existence proof failure due to oversimplification of set properties. \\
\textcolor{red}{$\times$}~Assuming a simple `building block'' approach for unique representation. \quad
\textcolor{red}{$\times$}~Ignoring structural constraints that prevent closure. \\
\textcolor{teal}{$\checkmark$}~Verify closure properties; test uniqueness implications on concrete examples. \\
\textit{Correct approach:} Recognise that uniqueness constraints lead to unavoidable contradictions.\\

\textbf{Example 2.} \dots \\

\textbf{Example 3.} \textit{Set Theory and Number Theory} --- Incorrect assumptions about consecutive pairs in infinite sets. \\
\textcolor{red}{$\times$}~Transferring properties of special numbers to arbitrary set elements. \quad
\textcolor{red}{$\times$}~Flawed casework that does not exploit the defining relationship. \\
\textcolor{teal}{$\checkmark$}~Ensure arguments rely only on defining relationships, not assumed structure. \\
\textit{Correct approach:} Assume both sets have finitely many consecutive pairs and derive a contradiction. \\[2pt]

\midrule
\rowcolor{gray!10}
\textbf{Response with RAG} \hfill \textcolor{teal}{7/8} \\
\mycomment{Based on the provided examples, the model recognizes that the number of cousins factors multiplicatively over independent arithmetic-progression blocks:}
if $S$ is a union of progressions of sizes $k_1,\ldots,k_s$, then $N_C(S)=(k_1{+}1)(k_2{+}1)\cdots(k_s{+}1)$. Setting this equal to $4040 = 2^3\cdot 5\cdot 101$ and minimising $\sum k_i$ yields factors $(k_j{+}1)\in\{2,2,2,5,101\}$, giving $n=1+1+1+4+100=\boxed{\mathbf{107}}$. \\

\bottomrule
\end{tabular}
}

\end{minipage}
}
\caption{\textbf{A case study of \method- \tagRefl.} Without retrieval, Gemini-2.5-Flash fails to reach a correct answer in 8 attempts. Retrieval over full traces is also insufficient and does not lead to a correct solution. In contrast, retrieval over \method\ provides targeted reasoning guidance that enables the model to solve 7 out of 8 attempts correctly. Retrieved examples and solutions are shortened for brevity. Our comments on model behavior are in \texttt{\mycomment{dark blue}.}}
\vspace{-1em}

\label{fig:case-study}
\end{figure*}
We study how reasoning trajectories can serve as effective retrieval units for reasoning-intensive tasks. The key idea is to view trajectory retrieval as a representation problem, where the same trace can be transformed into different retrieval-friendly forms.

\subsection{Thinking Trajectory-Based Corpus Design}
\label{sec:method_setup}

Let $q \in \mathcal{Q}$ denote a test query and let $L$ be the target model used for inference. We assume access to an auxiliary collection of problems and their associated reasoning trajectories, from which we construct a set of reasoning trajectories
$
\mathcal{T} = \{\tau_1, \tau_2, \dots, \tau_n\},
$
where each $\tau_i$ is a raw reasoning trace generated for an auxiliary problem by a strong model. These traces form the starting point of our corpus construction pipeline.

From this set of trajectories, we derive a trajectory-based retrieval corpus $\mathcal{C}_{\tau}$, where each retrieval unit corresponds either to a full or chunked raw trajectory $\tau_i \in \mathcal{T}$ . Given $\mathcal{C}_{\tau}$, a retriever $R$ returns the top-$k$ units
$D(q; \mathcal{C}_{\tau}, k) = \{\tau_1, \dots, \tau_k\}.$
The retrieved units are then concatenated ($\oplus$) with the query and provided to the model:
$y \sim L(D(q; \mathcal{C}_{\tau}, k) \oplus q).$
We study whether thinking traces are an effective retrieval corpus for reasoning-intensive RAG, and whether transforming them into more structured forms can make them even more useful.
\subsection{T3: Transformation of Thinking Traces}
\label{sec:method_reconstruction}
We model reasoning transformation as a family of offline functions applied to a corpus of raw reasoning trajectories. Each transformation maps a trajectory $\tau$ into one or more retrieval-oriented representations,
$
f : \tau \mapsto \{\tilde{\tau}_1, \dots, \tilde{\tau}_m\}.
$
Applying such a transformation to the full trajectory set $\mathcal{T}$ yields a transformed trajectory corpus
$\tilde{\mathcal{C}}_{\tau} =\bigcup\limits_{\tau \in \mathcal{T} }f(\tau)$.

In general, each raw trajectory may produce one or more transformed representations, and these transformed units are often shorter than the original trajectory, i.e., $|\tilde{\tau}_i| \ll |\tau|$, reflecting different degrees of compression and abstraction.

This formulation has two advantages. First, all transformations are \emph{query-independent} and can therefore be applied fully offline, incurring only a one-time cost while enabling reuse of the transformed corpus across future queries. Second, because all retrieval variants are derived from the same base trajectory set, we can isolate the effect of representation design while keeping the retrieval and generation pipeline fixed.

We present three query-independent strategies for reconstructing raw reasoning traces, each one capturing a distinct perspective on what to preserve from the original trajectory. Prompts for each transformation are available in Appendix \ref{sec:app:prompts}.
\paragraph{Structural Normalization \tagStruc}
This strategy rewrites raw traces into concise step-by-step procedures. It removes detours, noise, and inconsistent formatting of raw thinking traces while preserving the main reasoning structure, producing cleaner scaffolds that are easier to retrieve and reuse as inference-time guidance.

\paragraph{Semantic Distillation \tagSem.}
This strategy keeps the core idea of traces while removing lower-level details. Rather than preserving every intermediate step, it emphasizes the key decisions and central insight of the solution. This lets us test whether retrieval benefits more from concise semantic guidance than from fuller procedural detail.

\paragraph{Reflection \tagRefl.}
This strategy rewrites a trace into a contrastive form of likely mistakes and how to avoid them. It highlights common errors, misleading intuitions, and critical checks, along with a brief statement of the right approach. This provides diagnostic context that can help the model avoid unproductive paths and recover from likely errors.

Figure~\ref{fig:case-study} shows an example where Gemini-2.5-Flash fails in both the LLM-only (No RAG) setup and with RAG over full traces, but succeeds once the retrieved traces are transformed with \tagRefl. This illustrates that, while raw or unstructured retrieved reasoning may be insufficient and noisy, transformed traces can provide the right guidance to help the model complete the solution.

\section{Experimental Setup}
\label{sec:experiments}
\subsection{Thinking Trace Sources}
\label{sec:tts}
We construct thinking-trajectory corpora from different LLMs and problem collections. We focus on a \textit{shared-corpus} setting, where previously generated traces are reused across inference models, allowing us to test the  \textit{transferability} of reasoning traces i.e., whether traces from one ``thinker'' can guide a different ``solver''. We consider two large-scale sources of reasoning-intensive questions; additional details are provided in Appendix~\ref{app:datasource}:

\noindent \textbf{\method-59k:} A collection of 59K reasoning-intensive problems paired with {Gemini-2-thinking} traces  , drawn primarily from math (53k) with additional science and general reasoning domains. We use the S1 data pipeline to curate this collection \citep{muennighoff2025s1simpletesttimescaling}.

\noindent  \textbf{\method-114K:} A collection of 114K reasoning problems paired with QwQ-32B thinking traces. It spans mathematics (89K), code (20K), science (4K), and puzzles (1K), and is obtained from the OpenThoughts data recipe \citep{guha2025openthoughtsdatarecipesreasoning}.

\textbf{Decontamination.} 
We take several steps to reduce contamination. First, we evaluate on recent benchmarks, including AIME 2025--2026, whose problems were released after the source problem collections used to construct our trace corpora. Second, following prior work \citep{borgeaud2022improving,lyu2025frustratinglysimpleretrievalimproves}, we decontaminate both trace collections against the evaluation benchmarks by removing samples whose similarity to an evaluation query exceeds a 13-gram Jaccard threshold. This removes approximately 1.8\% of the data. In Appendix~\ref{app:decontamination}, we further analyze the choice of this threshold and provide additional lexical and semantic contamination checks.

\textbf{Corpus diversity.} Since our corpus is built from many distinct auxiliary problems rather than repeated passages, we verify that this does not produce a repetitive or redundant retrieval corpus. We measure lexical and semantic overlap across 5,000 sampled passages from each corpus variant and find that near-duplicate passages (cosine similarity $>0.9$) are effectively absent (0.00\% across all variants), indicating that both raw and transformed traces remain diverse rather than collapsing into repeated templates. Full results are reported in Appendix~\ref{app:diversity}.

\renewcommand{\pos}[1]{\textcolor[HTML]{2e7d32}{\scriptsize(+#1)}}
\renewcommand{\neg}[1]{\textcolor[HTML]{c62828}{\scriptsize(#1)}}
\renewcommand{\neu}[1]{\textcolor[HTML]{555555}{\scriptsize(#1)}}
\newcommand{\nodata}{\multicolumn{1}{c}{--}}
\newcommand{\nodelta}{\scriptsize(--)}

\definecolor{topcell}{HTML}{D4EDDA}   
\definecolor{botcell}{HTML}{F8D7DA}   
\newcommand{\hi}[1]{\cellcolor{topcell}#1}
\newcommand{\lo}[1]{\cellcolor{botcell}#1}

\newcolumntype{A}{>{\centering\arraybackslash\large}m{0.052\linewidth}}
\newcolumntype{D}{>{\centering\arraybackslash}m{0.07\linewidth}}

\begin{table*}[t]
\centering
\resizebox{\linewidth}{!}{%
\renewcommand{\arraystretch}{1.15}
\setlength{\tabcolsep}{4.5pt}
\begin{tabular}{l
  A D A D A D @{\hspace{8pt}}
  A D A D A D @{\hspace{8pt}}
  A D A D A D
}
\toprule
& \multicolumn{6}{c}{\textbf{AIME 2025--2026}}
& \multicolumn{6}{c}{\textbf{GPQA-Diamond}}
& \multicolumn{6}{c}{\textbf{LiveCodeBench}} \\
\cmidrule(lr){2-7}\cmidrule(lr){8-13}\cmidrule(l){14-19}
& \multicolumn{2}{c}{\makecell{GPT-5}}
& \multicolumn{2}{c}{\makecell{GPT-OSS\\120B}}
& \multicolumn{2}{c}{\makecell{Gemini-2.5\\Flash}}
& \multicolumn{2}{c}{\makecell{GPT-5}}
& \multicolumn{2}{c}{\makecell{GPT-OSS\\120B}}
& \multicolumn{2}{c}{\makecell{Gemini-2.5\\Flash}}
& \multicolumn{2}{c}{\makecell{GPT-5}}
& \multicolumn{2}{c}{\makecell{GPT-OSS\\120B}}
& \multicolumn{2}{c}{\makecell{Gemini-2.5\\Flash}} \\
\cmidrule(lr){2-3}\cmidrule(lr){4-5}\cmidrule(lr){6-7}
\cmidrule(lr){8-9}\cmidrule(lr){10-11}\cmidrule(lr){12-13}
\cmidrule(lr){14-15}\cmidrule(lr){16-17}\cmidrule(l){18-19}
\textbf{Corpus}
  & Acc. & $\Delta\%$ & Acc. & $\Delta\%$ & Acc. & $\Delta\%$
  & Acc. & $\Delta\%$ & Acc. & $\Delta\%$ & Acc. & $\Delta\%$
  & Acc. & $\Delta\%$ & Acc. & $\Delta\%$ & Acc. & $\Delta\%$ \\
\midrule
\rowcolor{gray!8}
\multicolumn{19}{l}{\textit{Baseline}} \\
No RAG
  & \lo{86.7} & & 78.3 & & 53.3 &
  & 83.8 & & 70.7 & & 77.3 &
  & \lo{57.4} & & 57.9 & & 45.1 & \\
\midrule
\rowcolor{gray!8}
\multicolumn{19}{l}{\textit{General-purpose corpora}} \\
Random 
  & \lo{86.7} & \neu{0.0\%} & \lo{73.3} & \neg{-6.4\%} & 53.3 & \neu{0.0\%}
  & \lo{83.3} & \neg{-0.6\%} & 70.7 & \neu{0.0\%} & \lo{76.3} & \neg{-1.3\%}
  & 58.4 & \pos{1.7\%} & \lo{55.0} & \neg{-5.0\%} & 46.0 & \pos{2.0\%} \\
  
OpenWebMath
  & \lo{85.0} & \neg{-2.0\%} & \lo{63.3} & \neg{-19.2\%} & \lo{45.0} & \neg{-15.6\%}
  & \lo{82.8} & \neg{-1.2\%} & 69.2 & \neg{-2.1\%} & \lo{75.8} & \neg{-1.9\%}
  & 58.9 & \pos{2.6\%} & \lo{37.6} & \neg{-35.1\%} & \lo{39.6} & \neg{-12.2\%} \\
StackExchange
  & \lo{83.3} & \neg{-3.9\%} & 76.7 & \neg{-2.0\%} & \lo{46.7} & \neg{-12.4\%}
  & \lo{83.3} & \neg{-0.6\%} & \lo{34.3} & \neg{-51.5\%} & \lo{76.8} & \neg{-0.6\%}
  & \lo{57.4} & \neu{0.0\%} & \lo{55.0} & \neg{-5.0\%} & \lo{42.1} & \neg{-6.7\%} \\
Wikipedia-DPR
  & 88.3 & \pos{1.8\%} & \lo{71.7} & \neg{-8.4\%} & 56.7 & \pos{6.4\%}
  & 84.3 & \pos{0.6\%} & \hi{71.7} & \pos{1.4\%} & \hi{80.3} & \pos{3.9\%}
  & \hi{59.9} & \pos{4.4\%} & 59.4 & \pos{2.6\%} & 46.5 & \pos{3.1\%} \\
Wikipedia-RPJ
  & \hi{90.0} & \pos{3.8\%} & 76.7 & \neg{-2.0\%} & \lo{41.7} & \neg{-21.8\%}
  & \hi{85.9} & \pos{2.5\%} & \hi{72.2} & \pos{2.1\%} & 78.3 & \pos{1.3\%}
  & 58.9 & \pos{2.6\%} & 58.9 & \pos{1.7\%} & \lo{45.0} & \neg{-0.2\%} \\
GitHub
  & \hi{91.7} & \pos{5.8\%} & 76.7 & \neg{-2.0\%} & 60.0 & \pos{12.6\%}
  & 84.8 & \pos{1.2\%} & 68.2 & \neg{-3.5\%} & \hi{80.8} & \pos{4.5\%}
  & \lo{56.9} & \neg{-0.9\%} & \lo{54.5} & \neg{-5.9\%} & \lo{42.1} & \neg{-6.7\%} \\
Arxiv
  & \lo{85.0} & \neg{-2.0\%} & 78.3 & \neu{0.0\%} & 51.7 & \neg{-3.0\%}
  & 84.8 & \pos{1.2\%} & 69.7 & \neg{-1.4\%} & 77.3 & \neu{0.0\%}
  & \lo{57.9} & \pos{0.9\%} & 57.4 & \neg{-0.9\%} & 46.9 & \pos{4.0\%} \\
CompactDS
  & 88.3 & \pos{1.8\%} & 80.0 & \pos{2.2\%} & 58.3 & \pos{9.4\%}
  & \lo{82.8} & \neg{-1.2\%} & \lo{67.7} & \neg{-4.2\%} & 77.3 & \neu{0.0\%}
  & \hi{60.9} & \pos{6.1\%} & 57.4 & \neg{-0.9\%} & 45.4 & \pos{0.7\%} \\
Tavily Search API
  & \lo{83.3} & \neg{-3.9\%} & 75.0 & \neg{-4.2\%} & 60.0 & \pos{12.6\%}
  & 84.8 & \pos{1.2\%} & \lo{59.6} & \neg{-15.7\%} & \hi{79.8} & \pos{3.2\%}
  & 58.4 & \pos{1.7\%} & \hi{59.9} & \pos{3.5\%} & \hi{47.9} & \pos{6.2\%} \\
\midrule
\rowcolor{gray!8}
\multicolumn{19}{l}{\textit{Thinking traces-based corpora}} \\
Full traj.
  & \lo{86.7} & \neu{0.0\%} & \lo{73.3} & \neg{-6.4\%} & 73.3 & \pos{37.5\%}
  & \lo{80.8} & \neg{-3.6\%} & 69.2 & \neg{-2.1\%} & \lo{76.3} & \neg{-1.3\%}
  & \lo{57.9} & \pos{0.9\%} & 58.9 & \pos{1.7\%} & 46.5 & \pos{3.1\%} \\
Chunked traj.
  & \hi{91.7} & \pos{5.8\%} & \hi{\textbf{85.0}} & \textbf{\pos{8.6\%}} & \hi{80.0} & \pos{50.1\%}
  & 84.8 & \pos{1.2\%} & \hi{71.7} & \pos{1.4\%} & 79.3 & \pos{2.6\%}
  & \hi{\textbf{60.9}} & \textbf{\pos{6.1\%}} & 58.9 & \pos{1.7\%} & \hi{\textbf{48.0}} & \textbf{\pos{6.4\%}} \\
\method-59k-\tagStruc
  & \hi{91.7} & \pos{5.8\%} & \hi{81.7} & \pos{4.3\%} & 73.3 & \pos{37.5\%}
  & \hi{\textbf{87.4}} & \textbf{\pos{4.3\%}} & 70.7 & \neu{0.0\%} & \hi{\textbf{80.8}} & \textbf{\pos{4.5\%}}
  & \hi{60.4} & \pos{5.2\%} & \hi{\textbf{61.4}} & \textbf{\pos{6.0\%}} & \hi{47.0} & \pos{4.2\%} \\
\method-59k-\tagRefl
  & \hi{\textbf{93.3}} & \textbf{\pos{7.6\%}} & \hi{81.7} & \pos{4.3\%} & \hi{76.7} & \pos{43.9\%}
  & 84.3 & \pos{0.6\%} & \hi{71.7} & \pos{1.4\%} & 79.3 & \pos{2.6\%}
  & \hi{59.9} & \pos{4.4\%} & 58.4 & \pos{0.9\%} & \lo{45.0} & \neg{-0.2\%} \\
\method-59k-\tagSem
  & 88.3 & \pos{1.8\%} & \hi{83.3} & \pos{6.4\%} & \hi{\textbf{83.3}} & \textbf{\pos{56.3\%}}
  & \hi{86.4} & \pos{3.1\%} & \hi{\textbf{74.7}} & \textbf{\pos{5.7\%}} & 78.8 & \pos{1.9\%}
  & 58.9 & \pos{2.6\%} & \hi{60.9} & \pos{5.2\%} & \lo{45.0} & \neg{-0.2\%} \\
\bottomrule
\end{tabular}}
\caption{
Results on AIME 2025--2026, GPQA-Diamond, and LiveCodeBench.
The best score in each column is shown in bold; \colorbox{topcell}{\strut green} cells mark the three highest accuracies per column and \colorbox{botcell}{\strut red} cells the three lowest (ties included). The relative improvement over the No RAG baseline is reported.}
\vspace{-1em}
\label{tab:rag-main}
\end{table*}

\textbf{Transformation.}
All transformed variants are generated by applying the prompts in Appendix~\ref{sec:app:prompts} with Gemini-2-Flash-Lite. We use a smaller model because transformation is applied once to the entire corpus. On the 59K corpus, each transformation costs at most \$26.74, and all three transformed variants cost only \$64.74 in total. Even with a substantially more expensive model, this remains a modest one-time preprocessing cost. Unlike trace generation, which requires expensive long-form reasoning from strong models, transformation is a lightweight rewrite of existing traces. This makes it practical to construct shared reasoning corpora that can be reused across inference models. We provide a cost and token breakdown in Appendix~\ref{app:transformation_cost}.
\subsection{Inference Setup}
We consider a diverse set of frontier LLMs, GPT-5, GPT-OSS-120B, and Gemini-2.5-Flash, deliberately spanning different scales, reasoning capabilities, and open- and closed-source families to study how retrieval interacts with different deployment regimes and model generations while still focusing on strong contemporary reasoners. Since these models are already strong reasoners, we separately evaluate whether \method{} also benefits substantially smaller, less capable target models in Appendix~\ref{app:smaller_models}.

For retrieval, we use \texttt{e5-base-v2} as our primary encoder for both queries and thinking traces, retrieving the top-3 documents by default. For each problem, we augment the original question with the retrieved examples, formatted as a hint-augmented prompt shown in Figure~\ref{fig:rag-reasoning-prompt}. We compare retrieval over full trajectories, which treat each thinking trace as a single retrieval unit, with chunked trajectories, where traces are split into fixed-length 512-token segments. For transformed traces, we use 2048-token retrieval units and do not apply additional chunking, since they are already substantially shorter on average, typically below 1000 words. Further analysis of trace lengths is provided in Appendix~\ref{app:datasource}.
We also provide ablations on the number of retrieved documents and on using Qwen3-Embedding-0.6B in Appendix~\ref{app:retrieval}. These ablations follow the same pattern as our main results i.e., top-3 retrieval is a stable default, and the choice of encoder does not change our main conclusions. Since we retrieve $k{=}3$ documents per query, Appendix~\ref{app:retrieval} also audits whether the retrieved traces conflict with one another; we find that retrieved sets are overwhelmingly complementary or reinforcing rather than contradictory.

\subsection{Baselines}
We compare against No RAG LLM-only setup, and RAG over several general-purpose corpora. These include major CompactDS subsets~\citep{lyu2025frustratinglysimpleretrievalimproves}: OpenWebMath (6.4M documents), StackExchange (29.8M), Wikipedia-DPR (21.0M), Wikipedia-RPJ (29.8M), GitHub (28.8M), and ArXiv (1.6M). All are chunked into 512-token passages and indexed with the same \texttt{e5-base-v2} retriever.  We also include random retrieval, where passages are sampled uniformly across all general-purpose corpora, to verify that gains come from relevant retrieval.
Indexing the full CompactDS corpus (639M+ documents) with our setup is computationally expensive, so we additionally report DS-Serve results~\citep{liu2026ds}, which serve the full corpus with Contriever~\citep{izacard2021unsupervised}. We also include Tavily Search API\footnote{\url{https://www.tavily.com/}} as a commercial real-time web search baseline. All corpora and retrieved results are decontaminated with respect to the evaluation benchmarks as described in Section~\ref{sec:tts}.

\subsection{Evaluation}

\oldparagraph{Benchmarks.}
We evaluate on a diverse set of reasoning benchmarks including 
1) {AIME (2025--2026):} Competition-level math problems, where each year consists of 30  questions.
2) GPQA-Diamond: A benchmark of 198 graduate-level scientific questions across biology, chemistry, and physics \cite{rein2023gpqagraduatelevelgoogleproofqa}.
3) {LiveCodeBench}: A subset of 202 programming problems constructed from LCB  V3 and V4 \citep{jain2024livecodebenchholisticcontaminationfree}.

AIME 2025--2026 is challenging even for frontier models, which makes it difficult to evaluate substantially smaller models whose baseline accuracy is near floor. For the smaller-model evaluation in Appendix~\ref{app:smaller_models}, we therefore additionally report results on MATH~\citep{hendrycks2021measuring}, a reasoning-intensive benchmark better calibrated to weaker models' capabilities.

\paragraph{Evaluation.}
We evaluate all pipelines using EleutherAI LM Evaluation Harness \citep{eval-harness} with custom task definitions. 
We query the target model through an OpenRouter-hosted OpenAI-compatible interface. Unless otherwise specified, we allow up to 16K generation tokens and use a temperature of 0.6 when applicable.

To reduce variance from stochastic generation, we sample multiple independent responses per problem. For AIME, we use 8 samples per query and report Average@8. For the larger GPQA-Diamond and LiveCodeBench benchmarks, we use 4 samples per query and report Average@4.

Answers are automatically extracted from model outputs and scored against the gold solution. For AIME and GPQA-Diamond, we report exact-match accuracy. For LiveCodeBench, each sampled program is evaluated using the standard pass@1 criterion, and the reported score is averaged over 4 samples. When simple parsing is insufficient, we use GPT-4o-mini only for answer normalization during post-processing.

\section{Results}
\label{sec:results}

\begin{table*}[t!]
\centering
\label{tab:ablation}
\resizebox{\linewidth}{!}{%
\renewcommand{\arraystretch}{1.15}
\setlength{\tabcolsep}{4.5pt}
\begin{tabular}{l
  A D A D A D @{\hspace{8pt}}
  A D A D A D @{\hspace{8pt}}
  A D A D A D
}
\toprule
& \multicolumn{6}{c}{\textbf{AIME 2025--2026}}
& \multicolumn{6}{c}{\textbf{GPQA-Diamond}}
& \multicolumn{6}{c}{\textbf{LiveCodeBench}} \\
\cmidrule(lr){2-7}\cmidrule(lr){8-13}\cmidrule(l){14-19}
& \multicolumn{2}{c}{\makecell{GPT-5}}
& \multicolumn{2}{c}{\makecell{GPT-OSS\\120B}}
& \multicolumn{2}{c}{\makecell{Gemini-2.5\\Flash}}
& \multicolumn{2}{c}{\makecell{GPT-5}}
& \multicolumn{2}{c}{\makecell{GPT-OSS\\120B}}
& \multicolumn{2}{c}{\makecell{Gemini-2.5\\Flash}}
& \multicolumn{2}{c}{\makecell{GPT-5}}
& \multicolumn{2}{c}{\makecell{GPT-OSS\\120B}}
& \multicolumn{2}{c}{\makecell{Gemini-2.5\\Flash}} \\
\cmidrule(lr){2-3}\cmidrule(lr){4-5}\cmidrule(lr){6-7}
\cmidrule(lr){8-9}\cmidrule(lr){10-11}\cmidrule(lr){12-13}
\cmidrule(lr){14-15}\cmidrule(lr){16-17}\cmidrule(l){18-19}
\textbf{Corpus}
  & Acc. & $\Delta\%$ & Acc. & $\Delta\%$ & Acc. & $\Delta\%$
  & Acc. & $\Delta\%$ & Acc. & $\Delta\%$ & Acc. & $\Delta\%$
  & Acc. & $\Delta\%$ & Acc. & $\Delta\%$ & Acc. & $\Delta\%$ \\
\midrule
No RAG
  & 86.7 & & 78.3 & & 53.3 &
  & 83.8 & & 70.7 & & 77.3 &
  & 57.4 & & 57.9 & & 45.1 & \\
\midrule
Output
  & 81.7 & \neg{-5.8\%} & 83.3 & \pos{6.4\%} & 68.3 & \pos{28.1\%}
  & 84.8 & \pos{1.2\%} & 70.7 & \neu{0.0\%} & 80.8 & \pos{4.5\%}
  & 59.4 & \pos{3.5\%} & 58.4 & \pos{0.9\%} &\textbf{ 56.4} & \pos{25.1\%} \\
Thinking trajectories
  & \textbf{91.7} & \textbf{\pos{5.8\%}} & \textbf{85.0} & \textbf{\pos{8.6\%}} & \textbf{80.0} & \textbf{\pos{50.1\%}}
  & 84.8 & \pos{1.2\%} & \textbf{71.7} & \textbf{\pos{1.4\%}} & 79.3 & \pos{2.6\%}
  & \textbf{60.9} & \textbf{\pos{6.1\%}} & \textbf{58.9} & \textbf{\pos{1.7\%}} & 48.0 & \pos{6.4\%} \\
\midrule
QwQ-32B
  & 86.7 & \neu{0.0\%} & 78.3 & \neu{0.0\%} & 68.3 & \pos{28.1\%}
  & 83.3 & \neg{-0.6\%} & 71.2 & \pos{0.7\%} & 79.3 & \pos{2.6\%}
  & 57.4 & \neu{0.0\%} & \textbf{59.9} & \textbf{\pos{3.5\%}} & 44.1 & \neg{-2.2\%} \\
GPT-OSS-120B
  & 90.0 & \pos{3.8\%} & 80.0 & \pos{2.2\%} & 45.0 & \neg{-15.6\%}
  & 52.0 & \neg{-37.9\%} & 70.2 & \neg{-0.7\%} & 77.8 & \pos{0.6\%}
  & 57.9 & \pos{0.9\%} & 55.9 & \neg{-3.5\%} & 42.1 & \neg{-6.7\%} \\
Gemini-2-thinking
  & \textbf{91.7} & \textbf{\pos{5.8\%}} & \textbf{85.0} & \textbf{\pos{8.6\%}} & \textbf{80.0} & \textbf{\pos{50.1\%}}
  & \textbf{84.8} & \textbf{\pos{1.2\%}} & \textbf{71.7} & \textbf{\pos{1.4\%}} & \textbf{79.3} & \textbf{\pos{2.6\%}}
  & \textbf{60.9} & \textbf{\pos{6.1\%}} & 58.9 & \pos{1.7\%} & \textbf{48.0} & \textbf{\pos{6.4\%}} \\
\bottomrule
\bottomrule

\end{tabular}}
\caption{
Ablations on the retrieval source for RAG on reasoning traces.
\textit{Top}: comparing retrieval over corpus of thinking trajectories against retrieval over the same model's final output corpus, showing that the gains stem from access to the reasoning process rather than from exposure to related problems alone.
\textit{Bottom}: comparing RAG thinking trajectories generated by three different thinker models over the same 59K-problem set, showing that trace quality from the thinker strongly affects downstream performance.
}
\vspace{-1em}
\label{tab:ablation}
\end{table*}

We study three research questions:
\begin{itemize}[leftmargin=*,nosep]
\item \textbf{RQ1:} Is RAG over thinking traces helpful for reasoning-intensive tasks?
\item \textbf{RQ2:} Can thinking traces be transformed to serve as more effective context for RAG?
\item \textbf{RQ3:} How does retrieval over thinking traces affect the inference cost?
\end{itemize}

\subsection{RQ1: Retrieval over Thinking Traces}
\label{sec:rq1}

We compare three experimental settings of No RAG, retrieval over general-purpose corpora, and retrieval over raw thinking traces in Table~\ref{tab:rag-main}.
The key pattern is that general-purpose retrieval is highly corpus- and task-dependent and it is inconsistent,  whereas retrieval over thinking traces provides a more reliable signal for reasoning tasks and improves performance, often substantially.

General-purpose corpora provide mixed results. Some sources help in isolated cases, but none improves all models on benchmarks. For example, OpenWebMath and StackExchange often hurt AIME performance, while Wikipedia, ArXiv, and CompactDS alternate depending on the model and task. This remains true even for CompactDS, despite its much larger scale (639M+ documents). Similarly, Tavily Search API, a real-time web retrieval baseline, does not yield consistent gains. 
Overall, these results suggest that the bottleneck is not simply whether the corpus is large or web-scale, but whether the retrieved content is aligned with the reasoning process required by the task.

In contrast, retrieval over thinking traces is substantially more effective. On AIME, Gemini-2.5-Flash improves from 53.3 to 73.3 (+37.5\%) with full traces, and further to 80.0 (+50.1\%) with simple chunking. These gains are much larger than those from general-purpose retrieval, even though the thinking-trace corpus contains only $\sim$59K traces, orders of magnitude fewer than the general-purpose corpora. The benefit is not limited to weaker models: GPT-5 improves from 86.7 to 91.7 (+5.8\%) with chunked traces, showing that reasoning-oriented retrieval remains useful for frontier models. While our trace corpus is heavily skewed toward mathematics (Appendix~\ref{app:datasource}), we still observe improvements on GPQA and LiveCodeBench, though the gains are more modest.

A consistent pattern is that chunked traces outperform full traces, suggesting that long raw traces are often too verbose and noisy to serve as effective retrieval units. This motivates transforming traces into compact, retrieval-friendly representations.

\oldparagraph{Retrieval on Thinking Traces vs Output.}
We next compare retrieval over thinking trajectories with retrieval over final outputs in Table~\ref{tab:ablation}. While output retrieval is often beneficial, full reasoning trajectories are generally stronger, suggesting that gains come not only from related problems but from intermediate reasoning. The advantage is clearest on AIME, where thinking traces outperform output-only retrieval across all models. On GPQA and LiveCodeBench, the gap is smaller and occasionally mixed; for Gemini-2.5-Flash, output attempts slightly outperform traces on both benchmarks. Overall, RAG on thinking traces provides richer context than final answers alone.

\paragraph{Impact of Thinking Traces.}
We further analyze the impact of the model generating the thinking traces in the bottom section of   Table~\ref{tab:ablation} in the Appendix. To isolate this effect, we generate traces over the same set of 59K problems from~\citep{muennighoff2025s1simpletesttimescaling} using three different thinkers: QwQ-32B, GPT-OSS-120B, and Gemini-2-thinking. 
As shown in Table \ref{tab:ablation}, we observe that the quality of the thinker strongly affects downstream performance. While traces from QwQ-32B and GPT-OSS-120B are often helpful, Gemini-2-thinking consistently produces the most effective retrieval corpus across benchmarks and models. Notably, this holds despite all traces being derived from the same problems, indicating that how the reasoning is expressed matters more than the underlying data itself. 

\subsection{RQ2: Transforming Thinking Traces}
\label{sec:rq2}

In RQ1, we found that chunked raw traces often outperform full trajectories. We now ask whether transforming those traces can produce even better retrieval corpora. Results in the last section of Table~\ref{tab:rag-main} show that RAG with \method{} consistently outperforms both raw-trace retrieval and general-purpose corpora. This indicates that not only the presence of reasoning traces, but also how they are represented, plays a critical role in their usefulness.

The impact of transformation is most pronounced on AIME 2025--2026. For example, using \method-59k, \tagRefl{} reaches 93.3 for GPT-5, outperforming both No RAG (86.7) and the best raw-trace baseline (91.7). For Gemini-2.5-Flash, \tagSem reaches 83.3, again improving over No RAG (53.3) and raw full-trace retrieval (73.3), while RAG over general-purpose corpora reaches at most 60.0 in this setting.

The best transformation depends on the task. On {GPQA-Diamond}, \tagStruc performs best for {GPT-5} and {Gemini-2.5-Flash}, reaching 87.4 and 80.8, while \tagSem performs best for {GPT-OSS-120B} at 74.7. On {LiveCodeBench}, transformed traces remain competitive and often outperform raw traces. Across tasks, they consistently outperform general-purpose retrieval, even when domain mismatch limits absolute gains.
Additionally, transformation gains are larger for weaker inference models. On AIME, Gemini-2.5-Flash improves by 56.3\%, from 53.3 to 83.3, with RAG on \method{}, while GPT-5 improves by 7.6\%, from 86.7 to 93.3. This suggests that transformation is especially valuable when models rely more on external reasoning signals.

\oldparagraph{Impact of the problem source.}
We further study the impact of the underlying problem collection in Table~\ref{tab:rag-thinker-ablation} in the Appendix \ref{app:problem_source}. Although \method-114K is derived from a larger and broader problem set, \method-59K often yields stronger downstream performance after transformation. This is especially clear on AIME and GPQA, where transformed traces from \method-59K outperform those from \method-114K across most models and transformation types. These results suggest that corpus size alone is not the main driver of reasoning-oriented retrieval; the alignment between the source problem set, its reasoning style, and the target benchmarks is  also important.

Interestingly, our shared-corpus setup also lets us test whether reasoning traces can transfer across models, even when they are produced by a different or older thinker. The answer is often yes: For example as shown in Table \ref{tab:ablation} Gemini-2.5-Flash benefits substantially from Gemini-2-thinking traces, and GPT-OSS-120B also benefits from QwQ-32B traces, showing that the value of reasoning traces can transfer across  model generations and families.

Overall, in response to RQ2, we find that \textbf{while raw traces are already useful, transforming them into cleaner, more compact, or more diagnostic representations often yields stronger performance across tasks and models.}

\begin{figure}[t]
    \centering
        \includegraphics[width=0.5\textwidth]{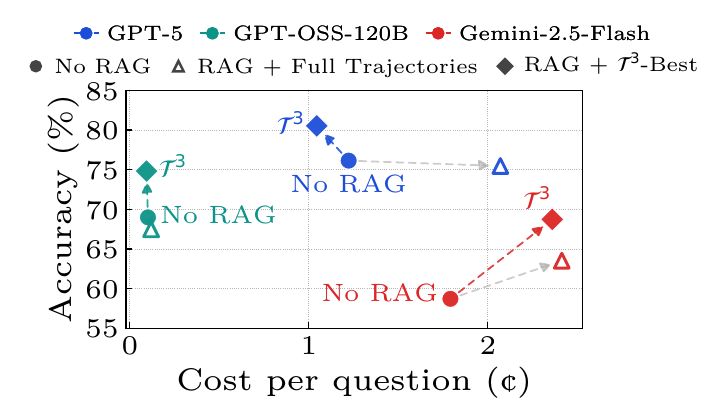}
\caption{
Average cost--accuracy trade-off across the three benchmarks. We compare No RAG, RAG over full thinking traces, and RAG with the best \method{} variant from Table~\ref{tab:rag-main}. Cost includes total tokens under each model's pricing; \method{} gives the strongest overall trade-off, often improving accuracy at lower or comparable cost.
}
\vspace{-1em}
    \label{fig:teaser}
\end{figure}
\subsection{Cost--Accuracy Trade-offs}
\label{sec:rq3}


Figure~\ref{fig:teaser} summarizes the average cost--accuracy trade-off across the three benchmarks. We report average inference cost per question, including both input and output tokens, together with average accuracy. We compare No RAG, RAG over full raw trajectories, and RAG with the best-performing \method{} variant.
A clear pattern is that full-trace retrieval is consistently the most expensive setting: it adds long contexts and often increases total inference cost. In contrast, \method{} provides a better cost--accuracy frontier by replacing full trajectories with shorter, more targeted transformed traces. For GPT-5, \method{} improves accuracy from 76.14 to 80.53 while reducing cost from 1.22 to 1.04 cents per query, a 14.8\% cost reduction relative to No RAG. For GPT-OSS-120B, accuracy improves from 68.99 to 74.82 while cost decreases slightly from 0.10 to 0.09 cents per query. For Gemini-2.5-Flash, \method{} increases cost relative to No RAG, but still achieves substantially higher accuracy and remains cheaper and more accurate than full-trace retrieval.

\textbf{Overall, \method{}  consistently dominates full thinking trace retrieval on the cost--accuracy frontier, and in some cases even improves reasoning at lower cost than No RAG. The effect is not universal, however: whether retrieved traces substitute for generation or stimulate more reasoning is strongly model-dependent.}

\section{Conclusion}
We revisit the role of retrieval in reasoning tasks and show that the limitation of RAG is not retrieval itself, but the choice of retrieval corpus. By shifting from retrieving documents to retrieving thinking traces, we demonstrate that even a simple retrieval-then-generate pipeline can significantly improve reasoning performance.
Our results show that raw thinking traces are already a strong retrieval source, and that transforming them with \method\ into more structured, compact, and diagnostic forms yields further gains across models and tasks. More broadly, we argue that thinking traces should be treated as a reusable resource that can be stored, transformed, and retrieved to support future reasoning.
\section*{Limitations}
\label{app:limit}

This work has several limitations. First, we study only vanilla RAG. This choice is intentional: our goal is to test whether simple retrieval over thinking traces can help reasoning in the first place. We leave more complex retrieval settings, such as iterative or reasoning-time retrieval, to future work.

Second, the trace corpora used in this work are heavily skewed toward mathematics. While we observe gains on GPQA-Diamond and LiveCodeBench, math, science, and code reasoning share procedural structure (e.g., decomposition, verification, backtracking), so this is a weak rather than a strict test of out-of-distribution generalization. We do not claim that \method{} generalizes to domains with little procedural overlap with the source corpus, and a cleaner test of true out-of-distribution transfer is left to future work.

Third, although we study traces from different strong reasoning models, we do not fully disentangle the effect of the thinker. In particular, we do not run a fully controlled comparison where the same auxiliary problem set is solved by many different thinker models and then evaluated with the same downstream solver. Such experiments would be informative, but are computationally expensive at the scale required here.

Despite these limitations, our results show that retrieval can be effective for reasoning when the right type of content is used. We hope this motivates future work on richer trace corpora, broader task coverage, and more adaptive retrieval mechanisms for reasoning-intensive RAG.

\section*{Acknowledgment}
AI assistants were used for editing and proofreading the manuscript. All scientific claims, experimental design, analyses, and final content were verified by the authors.

This work was supported in part by the Gemini Academic Program, Coefficient Giving, and compute provided by VESSL AI.

\bibliography{XX-references.bib}
\section*{Appendix}
\appendix
\section{Extended Related Work}
\label{app:related_work}

\oldparagraph{Reasoning and Retrieval.}
LLMs have shown strong performance on reasoning-intensive tasks such as mathematical problem solving, scientific question answering, and code generation \citep{wang2025surveylargelanguagemodels,rozière2024codellamaopenfoundation,auer2023sciqa}. Prior work improves reasoning through prompting strategies such as chain-of-thought and self-consistency \citep{wei2022chain,wang2022self}, distillation from stronger reasoning models \citep{ho2023largelanguagemodelsreasoning,magister2023teachingsmalllanguagemodels,shridhar2023distilling,muennighoff2025s1simpletesttimescaling}, and reinforcement learning with verifiable rewards \citep{guo2025deepseek,yu2025dapoopensourcellmreinforcement}. These approaches generally aim to improve the model's reasoning behavior directly, either through prompting, training, or distillation. Our work is complementary: we ask whether prior reasoning traces can instead be stored externally and reused through retrieval at inference time.

RAG has become a standard approach for improving LLMs on knowledge-intensive tasks by retrieving external documents that provide factual grounding and reduce hallucinations \citep{lewis2020retrieval,gao2023retrieval,siriwardhana2023improving}. Most work focuses on retrieving textual evidence from large corpora and improving how it is selected, structured, and incorporated into the model input \citep{fan2024survey,singal2024evidence,huo2023retrieving}. Recent work also studies retrieval from a scaling perspective: \citet{shao2024scalingretrievalbasedlanguagemodels} show that increasing datastore size can improve retrieval-based language models. However, for reasoning-intensive tasks, retrieved documents may provide relevant facts without exposing the reasoning process needed to use them. This motivates retrieval sources that contain process-level signals rather than only factual evidence.

\paragraph{RAG for Reasoning.}
A growing body of work explores retrieval in reasoning settings. \citet{levonian2023retrievalaugmentedgenerationimprovemath} show that retrieval can improve math question answering, but also introduces trade-offs between groundedness and human preference. RAT \citep{wang2024ratretrievalaugmentedthoughts} tightly couples retrieval with generation: it first produces an initial chain of thought and then revises each reasoning step using retrieved information. TRACE \citep{fang-etal-2024-trace} focuses on multi-hop QA and converts retrieved documents into knowledge-grounded reasoning chains using additional LLM calls before answer generation. These methods show that retrieval can help reasoning when retrieved information is structured or integrated into intermediate reasoning, but they still rely on online reasoning-chain construction or revision for each query.

\paragraph{Thought Reuse and Template-Based Reasoning.}
BoT \citep{yang2024buffer} and RoT \citep{ahmed2025retrievalofthoughtefficientreasoningreusing} are closest in spirit to our work because they also reuse prior reasoning artifacts. BoT maintains a meta-buffer of high-level thought templates distilled from previous problem-solving processes; for a new problem, it retrieves a relevant template and instantiates it through a dedicated reasoning procedure. RoT decomposes prior solutions into reusable reasoning steps, organizes them into a thought graph, and performs reward-guided traversal at inference time to assemble a problem-specific template. These methods demonstrate the value of reusable reasoning artifacts, but their retrieved objects are templates or graph nodes rather than native long-form thinking traces produced by reasoning models, and their use requires specialized inference-time mechanisms such as template instantiation, graph traversal, or thinking intervention.

\paragraph{Comparison to Our Work.}
Table~\ref{tab:related-work-compare} summarizes the main distinctions. Prior work improves reasoning RAG by modifying how retrieval is used at inference time: revising chains of thought, constructing evidence chains, instantiating templates, or traversing thought graphs. In contrast, our work studies the corpus-design question under a fixed retrieve-then-generate pipeline. We transform existing thinking traces offline into structured, compact, or diagnostic retrieval units, and then use them directly as context in a standard RAG setup.

\section{Smaller Target Models}
\label{app:smaller_models}

Our main experiments focus on frontier-class models (GPT-5, GPT-OSS-120B, Gemini-2.5-Flash), which are already strong reasoners with substantial reasoning-focused post-training. This leaves open whether \method{} also benefits less capable target models, which is arguably the more common deployment setting for retrieval-augmented reasoning in practice. We evaluate four smaller open models on MATH~\citep{hendrycks2021measuring}, a reasoning-intensive benchmark better calibrated to weaker models' baseline capabilities than AIME 2025--2026, where smaller models otherwise perform near floor.

Table~\ref{tab:math-smaller-models} reports Structural Normalization against No RAG and retrieval over raw chunked traces (i.e., the untransformed thinking traces split into fixed-length segments, described in Section~\ref{sec:experiments}) across four smaller open models on MATH. Raw-trace retrieval provides mixed gains relative to No RAG, sometimes underperforming it. In contrast, \tagStruc improves all four models, with gains of 6.8--13.6 points over No RAG and 5.4--16.1 points over raw chunked traces. This suggests that the benefit of \method{} is not restricted to frontier models that can simply follow a provided trace map that substantially smaller models can also interpret and apply retrieved reasoning guidance, and transforming the traces appears particularly important for these models, since raw traces alone are less reliably helpful.

\begin{table}[t]
\centering
\footnotesize
\setlength{\tabcolsep}{3pt}
\renewcommand{\arraystretch}{1.0}
\begin{tabular}{@{}lccc@{}}
\toprule
\textbf{Target Model} & \textbf{No RAG} & \textbf{Raw Traces} & \textbf{\tagStruc} \\
\midrule
Llama-3.1-8B           & 45.9 & 48.1 & 53.5 \\
Qwen2.5-32B             & 73.6 & 74.7 & 80.4 \\
OpenMath2-Llama3.1-8B   & 63.0 & 61.0 & 72.2 \\
Mathstral-7B-v0.1       & 47.5 & 45.0 & 61.1 \\
\bottomrule
\end{tabular}
\caption{
\method{} (Structural Normalization) vs.\ No RAG and retrieval over raw, chunked reasoning traces (i.e., unmodified traces split into fixed-length segments, as in Section~\ref{sec:experiments}) on MATH, across four smaller open models.
}
\label{tab:math-smaller-models}
\end{table}

\begin{figure*}[h!]
\vspace{1em}

\centering
\begin{tcolorbox}[title= \method{} -  Structural Normalization,
                  colframe=black!40, colback=gray!5, coltitle=black,
                  fonttitle=\bfseries, width=\linewidth, boxrule=0.5pt]
\small
\textbf{Instruction.} Convert the reasoning trace into a concise step-by-step cheatsheet.

\medskip
\textbf{Guidelines.}
\begin{itemize}[leftmargin=*,nosep]
  \item Use at most 7 steps.
  \item Each step should represent a meaningful action.
  \item Keep explanations short and clear.
  \item Focus on reusable reasoning patterns.
  \item Remove failed or irrelevant attempts.
\end{itemize}

\medskip
\textbf{Output format.}

\texttt{Problem: ...}

\texttt{Step 1: ...}\\
\texttt{...}\\
\texttt{Step N: ...}

\texttt{Answer: }\verb|$\boxed{[final answer]}$|

\medskip
\textbf{Given trace:} \verb|{trace}|
\end{tcolorbox}
\vspace{-1em}

\caption{Prompt for \tagStruc transformation.}
\label{fig:cheatsheet-prompt}
\end{figure*}

\begin{figure*}[t]
\centering

\begin{tcolorbox}[title= \method{} - Semantic Distillation,
                  colframe=black!40, colback=gray!5, coltitle=black,
                  fonttitle=\bfseries, width=\linewidth, boxrule=0.5pt]
\small
\textbf{Instruction.} Compress the reasoning trace into progressively more abstract representations.

\medskip
\textbf{Guidelines.}
\begin{itemize}[leftmargin=*,nosep]
  \item Provide three levels of abstraction.
  \item Pass 1: structured outline of reasoning steps.
  \item Pass 2: only key decisions or moves.
  \item Pass 3: a single core insight.
  \item Ensure consistency across all passes.
  \item Avoid redundancy.
\end{itemize}

\medskip
\textbf{Output format.}

\texttt{Problem: ...}

\texttt{Pass 1: ...}

\texttt{Pass 2: ...}

\texttt{Pass 3: ...}

\texttt{Answer: }\verb|$\boxed{[final answer]}$|

\medskip
\textbf{Given trace:} \verb|{trace}|
\end{tcolorbox}
\vspace{-1em}
\caption{Prompt for \tagSem transformation.}
\vspace{-1em}

\label{fig:multipass-prompt}
\end{figure*}

\vspace{2em}

\begin{figure*}[h!]
\centering
\vspace{1em}
\begin{tcolorbox}[title= \method{} - Reflection,
                  colframe=black!40, colback=gray!5, coltitle=black,
                  fonttitle=\bfseries, width=\linewidth, boxrule=0.5pt]
\small
\textbf{Instruction.} Extract failure patterns and negative knowledge from the reasoning trace.

\medskip
\textbf{Guidelines.}
\begin{itemize}[leftmargin=*,nosep]
  \item Focus on common mistakes and misleading reasoning paths.
  \item Explain why these mistakes are tempting.
  \item Highlight how to detect and avoid them.
  \item Provide contrast with the correct approach.
  \item Do not reproduce the full solution.
\end{itemize}

\medskip
\textbf{Output format.}

\texttt{Problem: ...}

\texttt{Common Mistakes: ...}

\texttt{Misleading Intuitions: ...}

\texttt{Critical Checks: ...}

\texttt{Correct Approach (brief): ...}

\texttt{Answer: }\verb|$\boxed{[final answer]}$|

\medskip

\textbf{Given trace:} \verb|{trace}|
\end{tcolorbox}
\vspace{-1em}
\caption{ Prompt for \tagRefl transformation.}
\label{fig:contrastive-prompt}
\end{figure*}

\begin{figure*}[h!]
\vspace{2em}
\centering
\begin{tcolorbox}[title=RAG Inference,
                  colframe=black!40, colback=gray!5, coltitle=black,
                  fonttitle=\bfseries, width=\linewidth, boxrule=0.5pt]
\small
\textbf{Instruction.} Solve the main problem by using useful hints and strategies from the retrieved examples.

\texttt{Example 1: ...}

\texttt{Example 2: ...}

\texttt{Example 3: ...}

\texttt{Main problem: ...}

\end{tcolorbox}
\caption{Prompt for RAG inference using retrieved examples.}
\label{fig:rag-reasoning-prompt}
\end{figure*}

\section{Prompts}\label{sec:app:prompts}

Here, we present the prompts used for transforming thinking traces and for RAG inference. The prompt used for transformation strategies introduced in Section \ref{sec:method_reconstruction}, namely Structural Normalization (\tagStruc), Semantic Distillation (\tagSem), and Reflection (\tagRefl) are shown in Figure \ref{fig:cheatsheet-prompt}, \ref{fig:multipass-prompt} and \ref{fig:contrastive-prompt}, respectively. Additionally, we provide our simple RAG inference prompt in Figure \ref{fig:rag-reasoning-prompt}. These prompts are applied to construct the transformed corpora and to guide the model at inference time.

\renewcommand{\pos}[1]{\textcolor[HTML]{2e7d32}{\scriptsize(+#1)}}
\renewcommand{\neg}[1]{\textcolor[HTML]{c62828}{\scriptsize(#1)}}
\renewcommand{\neu}[1]{\textcolor[HTML]{555555}{\scriptsize(#1)}}
\renewcommand{\nodata}{\multicolumn{1}{c}{--}}
\renewcommand{\nodelta}{\scriptsize(--)}

\begin{table*}[t]
\centering
 \resizebox{0.98\linewidth}{!}{%
\renewcommand{\arraystretch}{1.15}
\setlength{\tabcolsep}{5pt}
\begin{tabular}{c l
  A@{\hspace{1pt}}D A@{\hspace{1pt}}D A@{\hspace{1pt}}D @{\hspace{8pt}}
  A@{\hspace{1pt}}D A@{\hspace{1pt}}D A@{\hspace{1pt}}D @{\hspace{8pt}}
  A@{\hspace{1pt}}D A@{\hspace{1pt}}D A@{\hspace{1pt}}D
}
\toprule
& & \multicolumn{6}{c}{\textbf{AIME 2025--2026}}
& \multicolumn{6}{c}{\textbf{GPQA-Diamond}}
& \multicolumn{6}{c}{\textbf{LiveCodeBench}} \\
\cmidrule(lr){3-8}\cmidrule(lr){9-14}\cmidrule(l){15-20}
& & \multicolumn{2}{c}{\makecell{GPT-5}}
& \multicolumn{2}{c}{\makecell{GPT-OSS\\120B}}
& \multicolumn{2}{c}{\makecell{Gemini-2.5\\Flash}}
& \multicolumn{2}{c}{\makecell{GPT-5}}
& \multicolumn{2}{c}{\makecell{GPT-OSS\\120B}}
& \multicolumn{2}{c}{\makecell{Gemini-2.5\\Flash}}
& \multicolumn{2}{c}{\makecell{GPT-5}}
& \multicolumn{2}{c}{\makecell{GPT-OSS\\120B}}
& \multicolumn{2}{c}{\makecell{Gemini-2.5\\Flash}} \\
\cmidrule(lr){3-4}\cmidrule(lr){5-6}\cmidrule(lr){7-8}
\cmidrule(lr){9-10}\cmidrule(lr){11-12}\cmidrule(lr){13-14}
\cmidrule(lr){15-16}\cmidrule(lr){17-18}\cmidrule(l){19-20}
\textbf{} & \textbf{Method}
  & Acc. & $\Delta\%$ & Acc. & $\Delta\%$ & Acc. & $\Delta\%$
  & Acc. & $\Delta\%$ & Acc. & $\Delta\%$ & Acc. & $\Delta\%$
  & Acc. & $\Delta\%$ & Acc. & $\Delta\%$ & Acc. & $\Delta\%$ \\
\midrule
 & No RAG
  & 86.7 & & 78.3 & & 53.3 &
  & 83.8 & & 70.7 & & 77.3 &
  & 57.4 & & 57.9 & & 45.1 & \\
\midrule

\multirow{3}{*}{\rotatebox[origin=c]{90}{\textit{\method-114K}}}
& \tagStruc
  & 91.7 & \pos{5.8\%} & \textbf{88.3} & \textbf{\pos{12.8\%}} & 65.0 & \pos{22.0\%}
  & 81.3 & \neg{-3.0\%} & 69.7 & \neg{-1.4\%} & 76.8 & \neg{-0.6\%}
  & 59.4 & \pos{3.5\%} & 57.9 & \neu{0.0\%} & 46.0 & \pos{2.0\%} \\
& \tagRefl
  & 90.0 & \pos{3.8\%} & 80.0 & \pos{2.2\%} & 65.0 & \pos{22.0\%}
  & 84.3 & \pos{0.6\%} & 70.2 & \neg{-0.7\%} & 65.7 & \neg{-15.0\%}
  & 59.4 & \pos{3.5\%} & 55.4 & \neg{-4.3\%} & 44.6 & \neg{-1.1\%} \\
& \tagSem
  & 88.3 & \pos{1.8\%} & 80.0 & \pos{2.2\%} & 60.0 & \pos{12.6\%}
  & 84.8 & \pos{1.2\%} & 71.7 & \pos{1.4\%} & 58.1 & \neg{-24.8\%}
  & 59.4 & \pos{3.5\%} & 59.9 & \pos{3.5\%} & 43.6 & \neg{-3.3\%} \\
\midrule

\multirow{3}{*}{\rotatebox[origin=c]{90}{\textit{\method-59K}}}
& \tagStruc
  & 91.7 & \pos{5.8\%} & 81.7 & \pos{4.3\%} & 73.3 & \pos{37.5\%}
  & \textbf{87.4} & \textbf{\pos{4.3\%}} & 70.7 & \neu{0.0\%} & \textbf{80.8} & \textbf{\pos{4.5\%}}
  & \textbf{60.4} & \textbf{\pos{5.2\%}} & \textbf{61.4} & \textbf{\pos{6.0\%}} & \textbf{47.0} & \textbf{\pos{4.2\%}} \\
& \tagRefl
  & \textbf{93.3} & \textbf{\pos{7.6\%}} & 81.7 & \pos{4.3\%} & 76.7 & \pos{43.9\%}
  & 84.3 & \pos{0.6\%} & 71.7 & \pos{1.4\%} & 79.3 & \pos{2.6\%}
  & 59.9 & \pos{4.4\%} & 58.4 & \pos{0.9\%} & 45.0 & \neg{-0.2\%} \\
& \tagSem
  & 88.3 & \pos{1.8\%} & 83.3 & \pos{6.4\%} & \textbf{83.3} & \textbf{\pos{56.3\%}}
  & 86.4 & \pos{3.1\%} & \textbf{74.7} & \textbf{\pos{5.7\%}} & 78.8 & \pos{1.9\%}
  & 58.9 & \pos{2.6\%} & 60.9 & \pos{5.2\%} & 45.0 & \neg{-0.2\%} \\
\bottomrule
\end{tabular}}
\caption{
Ablation of \method{} on the problem source set.
Although \method-114K is derived from a larger and broader problem set (114K vs.\ 59K problems),
\method-59K often yields stronger downstream performance after transformation, suggesting that source-problem alignment and reasoning style can matter more than corpus size alone.
}
\label{tab:rag-thinker-ablation}
\end{table*}

\section{Thinking Traces}
\label{app:datasource}

\subsection{Corpora Statistics}
\begin{figure*}[h!]
    \centering
\vspace{-1em}
    \begin{subfigure}{\linewidth}
        \centering
        \includegraphics[width=\linewidth]{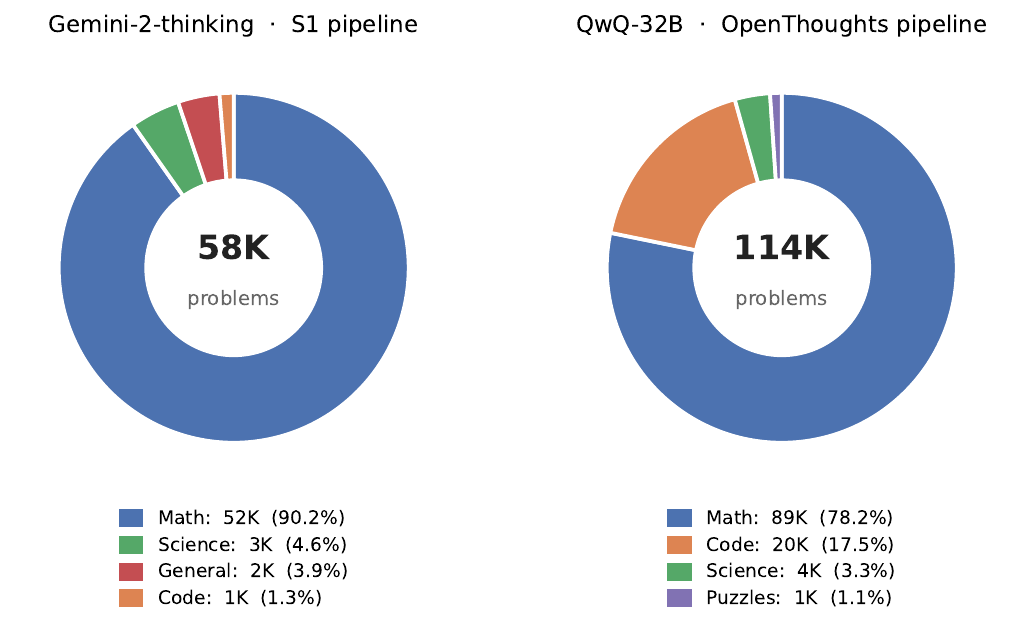}
        \caption{Domain distribution of the two thinking-trace corpora.}
        \label{fig:corpus-distribution}
    \end{subfigure}
    \hfill
    \begin{subfigure}{\linewidth}
        \centering
        \includegraphics[width=\linewidth, trim={0 0 0 1cm}, clip]{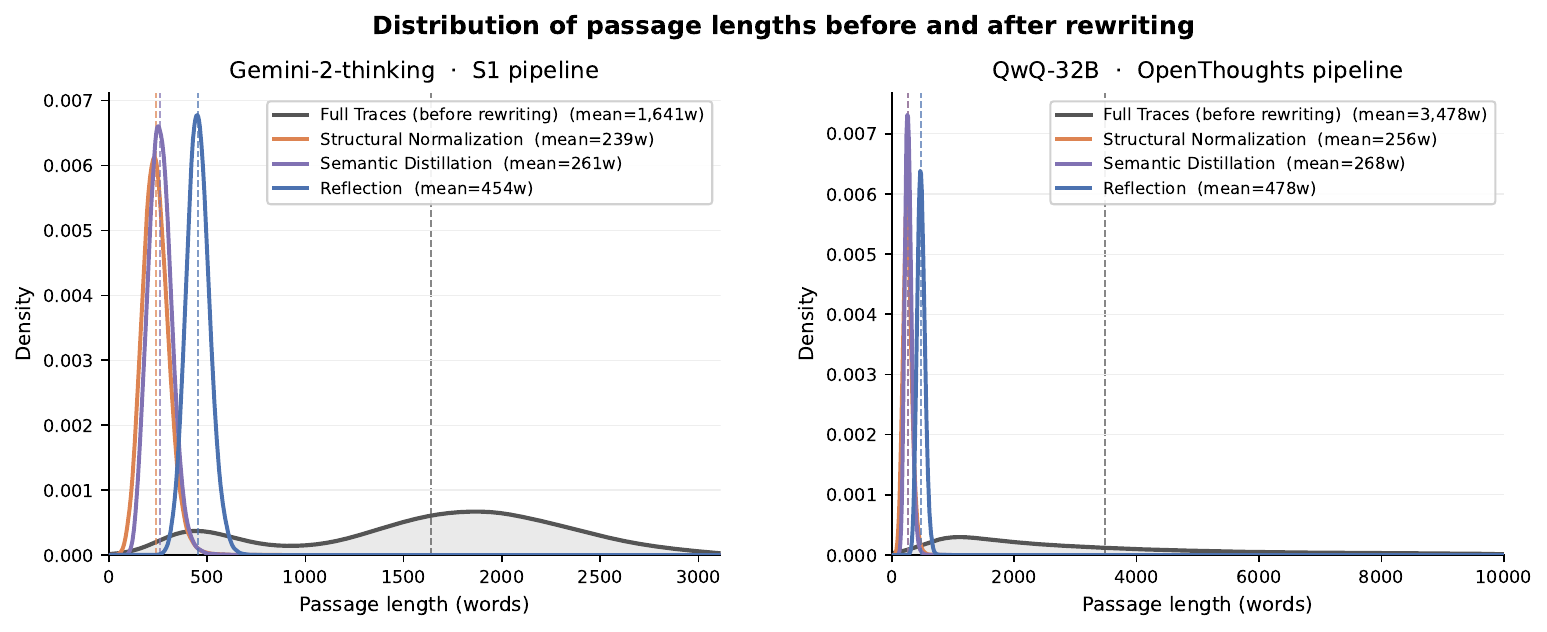}
        \caption{Passage length distribution before and after transformation.}
        \label{fig:trace_length}
    \end{subfigure}
    \caption{
    Corpus statistics for thinking traces. (Left) Domain distribution of the two corpora. Both are dominated by mathematical reasoning. Despite being smaller (58K after decontamination vs.\ 114K), \method-Gemini often yields stronger RAG performance, suggesting that trace quality may matter more than corpus size. (Right) Passage length distributions before and after transformation for both corpora. All transformed variants are substantially shorter than full traces, improving retrieval efficiency and reducing inference cost.
    }
    \vspace{-1em}
    \label{fig:corpus-stats}
\end{figure*}
\vspace{1em}

Here we show the domain distribution of the two thinking-trace corpora used in our experiments in Figure \ref{fig:corpus-distribution}. Both corpora are dominated by mathematical reasoning. \method-59k is more heavily skewed toward math (about 90\%) and contains very little code, whereas \method-114k includes a larger code component (17.5\%), reflecting its broader source coverage. Interestingly, despite being smaller and more narrowly focused, \method-59k often yields stronger RAG performance in our experiments. This suggests that trace quality may matter more than corpus size or breadth alone, potentially because the underlying Gemini-2-thinking model produces more useful reasoning traces for downstream retrieval.

We further analyze how transformation changes both the size and length of the resulting corpora. Structural normalization (\tagStruc) increases the number of passages by 35\% (78,522 vs.\ 58,071), because a single trajectory may be split into multiple procedural units when distinct steps or solution paths are extracted as separate documents. In contrast, semantic distillation (\tagSem) and reflection (\tagRefl) preserve the original number of trajectories, since each trace is rewritten into a single transformed representation.

We also examine the distribution of passage lengths before and after transformation. As shown in Figure~\ref{fig:trace_length}, full trajectories are substantially longer than all transformed variants. For the Gemini-based corpus, full traces have mean length 1,641 words, compared to 239 for structural normalization, 261 for semantic distillation, and 454 for reflection. The same pattern holds for the QwQ-based corpus, where full traces average 3,478 words, while the transformed variants average 256, 268, and 478 words, respectively. Overall, all three transformations produce much more compact retrieval units, which improves retrieval efficiency and reduces input cost at inference time.

\begin{table}[h!]
\centering
\footnotesize
\setlength{\tabcolsep}{3pt}
\renewcommand{\arraystretch}{1.2}
\begin{tabular}{@{}lrrrr@{}}
\toprule
\textbf{Trans.} & \textbf{Input Tok.} & \textbf{Output Tok.} & \textbf{Input \$} & \textbf{Output \$} \\
\midrule
\tagRefl  & 184.13M & 43.13M & \$13.81 & \$12.94 \\
\tagSem   & 156.93M & 22.73M & \$11.77 & \$6.82 \\
\tagStruc & 147.60M & 27.73M & \$11.07 & \$8.32 \\
\midrule
\textbf{Total} & 488.66M & 93.59M & \textbf{\$36.65} & \textbf{\$28.08} \\
\bottomrule
\end{tabular}
\caption{
Cost of generating transformed traces for the 59K corpus using Gemini-2.0-Flash-Lite.
\tagRefl corresponds to reflection, \tagSem to semantic distillation, and \tagStruc to structural normalization. The total cost is \$64.74.
}
\label{tab:transformation-cost}
\end{table}

\subsection{Problem Source Ablation}
\label{app:problem_source}

Table~\ref{tab:rag-thinker-ablation} compares transformed traces from two different problem sources: \method-114K, a larger and broader corpus, and \method-59K, a smaller corpus more concentrated on mathematical reasoning. Despite its smaller size, \method-59K often yields stronger downstream performance, particularly on AIME 2025--2026 and GPQA-Diamond. This suggests that, for reasoning-oriented retrieval, the usefulness of a trace corpus is not determined by scale alone. Source-problem alignment, reasoning style, and the structure of the generated traces can matter as much as, or more than, the number of available examples. Since the two corpora differ in both source distribution and trace-generation model, this ablation should be interpreted as a problem-source comparison rather than a controlled thinker-only comparison.

\subsection{Corpus Diversity}
\label{app:diversity}

Because each corpus is constructed by transforming or chunking a large number of distinct source trajectories rather than repeating a small set of passages, we verify that this process does not yield a highly redundant retrieval corpus. We sample 5,000 passages from each corpus variant and measure both lexical and semantic overlap: lexical overlap via trigram Jaccard similarity, and semantic overlap via \texttt{e5-base-v2} embedding cosine similarity. We report the near-duplicate rate as the fraction of sampled passage pairs with cosine similarity above 0.9.

Across all corpus variants, mean trigram Jaccard similarity remains below 0.04, and mean embedding cosine similarity ranges from 0.180 to 0.282 (Table~\ref{tab:corpus-diversity}). Critically, none of the sampled passage pairs exceeds a cosine similarity of 0.9, yielding a 0.00\% near-duplicate rate across all variants. This indicates that both raw and transformed traces remain lexically and semantically diverse: different auxiliary problems produce distinct passages rather than collapsing into a small number of repeated reasoning templates, even though transformation strategies such as \tagStruc and \tagRefl impose a shared structural or contrastive format on the output.

\subsection{Transformation Cost}
\label{app:transformation_cost}

We report the estimated cost of generating all transformed variants for the 59K trace corpus in Table~\ref{tab:transformation-cost}. Costs are computed using Gemini-2.0-Flash-Lite pricing of \$0.075 per million input tokens and \$0.30 per million output tokens. The three transformations together cost \$64.74 in total. Since this transformation is performed fully offline and only once per corpus, the resulting transformed traces can be reused across all downstream inference models and evaluation settings.

\subsection{Transformation Example}

Figures~\ref{fig:transform-example}, \ref{fig:transform-physics}, and \ref{fig:transform-coding2} present representative examples from math, physics, and coding domains, respectively. Each figure shows an example problem and its full reasoning trace (truncated for brevity), along with its transformed variants. As shown in these examples,   all transformed versions are significantly shorter compared to full traces, and each of them captures complementary aspects of the reasoning.

\section{Retrieval Ablation}
\label{app:retrieval}
\subsection{Impact of Number of Retrieved Documents}
\label{app:topk}
\begin{figure*}[t!]
    \centering

    \includegraphics[width=\linewidth]{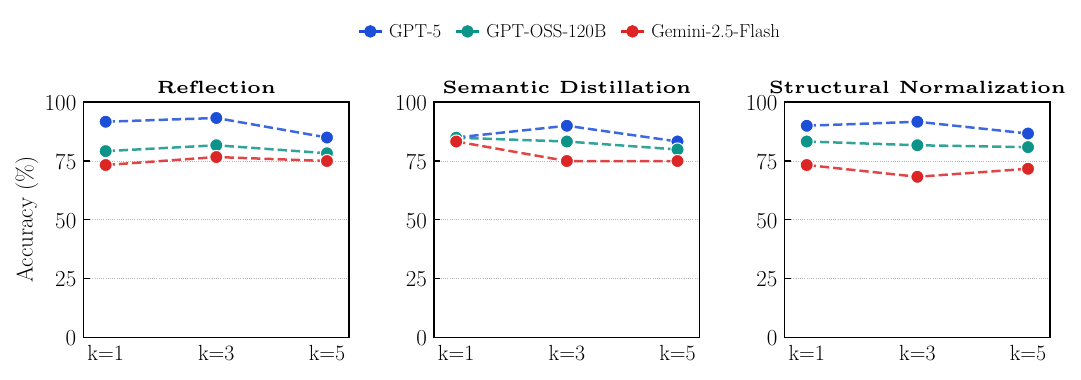}
    \caption{Impact of the number of retrieved documents ($k \in \{1, 3, 5\}$) on accuracy across three transformation strategies and three reader models on AIME 2025--2026. $k{=}3$ achieves the best or near-best accuracy most consistently across all settings.}

        \label{fig:topk}
\end{figure*}

We study how the number of retrieved documents affects downstream performance by evaluating top-k
 retrieval for $k \in \{1,3,5\}$ across all three \method{} transformations and all target models. As shown in Figure~\ref{fig:topk}, k=3 is the most consistent choice across models and methods. While k=1 or k=5 occasionally outperforms k=3 for a specific model–method combination,  k=3 yields the best or near-best accuracy on average. Retrieving a single document (k=1) can miss complementary reasoning steps, and k=5 sometimes introduces noisy or redundant context that degrades generation quality. We therefore use k=3 in all main experiments.

\begin{table}[t]
\centering
\footnotesize
\setlength{\tabcolsep}{2pt}
\renewcommand{\arraystretch}{1.0}
\begin{tabular}{@{}lccc@{}}
\toprule
\textbf{Corpus} & \makecell{\textbf{Trigram}\\\textbf{Jaccard}} & \makecell{\textbf{Cosine}\\\textbf{Sim.}} & \makecell{\textbf{Near-Dup.}\\\textbf{Rate}} \\
\midrule
Raw Traj.  & 0.005 & 0.180 & 0.00\% \\
\tagSem    & 0.024 & 0.216 & 0.00\% \\
\tagStruc  & 0.039 & 0.206 & 0.00\% \\
\tagRefl   & 0.028 & 0.282 & 0.00\% \\
\bottomrule
\end{tabular}
\caption{
Lexical (trigram Jaccard) and semantic (embedding cosine) similarity across 5,000 sampled passage pairs per corpus variant. Near-duplicate rate is the fraction of pairs with cosine similarity above 0.9. \tagStruc, \tagSem, and \tagRefl denote Structural Normalization, Semantic Distillation, and Reflection, respectively.
}
\label{tab:corpus-diversity}
\end{table}

 \subsection{Impact of Retriever}
 We evaluate whether our results depend on the choice of retriever by replacing \texttt{e5-base-v2} with Qwen3-Embedding-0.6B on Gemini-2-thinking traces. As shown in Table~\ref{tab:qwen3emb}, the overall pattern remains similar across retrievers. \texttt{e5-base-v2} is slightly better in several settings, but Qwen3-Embedding-0.6B also matches or improves over it in some cases, such as AIME with GPT-OSS-120B and GPQA-Diamond with Gemini-2.5-Flash. Importantly, the main conclusion is unchanged: transformed thinking traces consistently provide gains over No RAG across models and benchmarks. This suggests that \method{} is robust to the retriever choice rather than being tied to a single embedding model.

\begin{table*}[h!]
\centering
\resizebox{0.98\linewidth}{!}{%
\renewcommand{\arraystretch}{1.15}
\setlength{\tabcolsep}{4.5pt}
\begin{tabular}{l
  A D A D A D @{\hspace{8pt}}
  A D A D A D @{\hspace{8pt}}
  A D A D A D
}
\toprule
& \multicolumn{6}{c}{\textbf{AIME 2025--2026}}
& \multicolumn{6}{c}{\textbf{GPQA-Diamond}}
& \multicolumn{6}{c}{\textbf{LiveCodeBench}} \\
\cmidrule(lr){2-7}\cmidrule(lr){8-13}\cmidrule(l){14-19}
& \multicolumn{2}{c}{\makecell{GPT-5}}
& \multicolumn{2}{c}{\makecell{GPT-OSS\\120B}}
& \multicolumn{2}{c}{\makecell{Gemini-2.5\\Flash}}
& \multicolumn{2}{c}{\makecell{GPT-5}}
& \multicolumn{2}{c}{\makecell{GPT-OSS\\120B}}
& \multicolumn{2}{c}{\makecell{Gemini-2.5\\Flash}}
& \multicolumn{2}{c}{\makecell{GPT-5}}
& \multicolumn{2}{c}{\makecell{GPT-OSS\\120B}}
& \multicolumn{2}{c}{\makecell{Gemini-2.5\\Flash}} \\
\cmidrule(lr){2-3}\cmidrule(lr){4-5}\cmidrule(lr){6-7}
\cmidrule(lr){8-9}\cmidrule(lr){10-11}\cmidrule(lr){12-13}
\cmidrule(lr){14-15}\cmidrule(lr){16-17}\cmidrule(l){18-19}
\textbf{Corpus}
  & Acc. & $\Delta\%$ & Acc. & $\Delta\%$ & Acc. & $\Delta\%$
  & Acc. & $\Delta\%$ & Acc. & $\Delta\%$ & Acc. & $\Delta\%$
  & Acc. & $\Delta\%$ & Acc. & $\Delta\%$ & Acc. & $\Delta\%$ \\
\midrule
\rowcolor{gray!8}
\multicolumn{19}{l}{\textit{Baseline}} \\
No RAG
  & 86.7 & & 78.3 & & 53.3 &
  & 83.8 & & 70.7 & & 77.3 &
  & 57.4 & & 57.9 & & 45.1 & \\
\midrule
\rowcolor{gray!8}
\multicolumn{19}{l}{\textit{
e5-base-v2}} \\

\method- \tagStruc
  & {91.7} & \pos{5.8\%} & {81.7} & \pos{4.3\%} & 73.3 & \pos{37.5\%}
  & {\textbf{87.4}} & \textbf{\pos{4.3\%}} & 70.7 & \neu{0.0\%} & {\textbf{80.8}} & \textbf{\pos{4.5\%}}
  & \textbf{60.4} & \pos{5.2\%} & {\textbf{61.4}} & \textbf{\pos{6.0\%}} & \textbf{{47.0}} & \pos{4.2\%} \\
\method-\tagRefl
  & {\textbf{93.3}} & \textbf{\pos{7.6\%}} & {81.7} & \pos{4.3\%} & {76.7} & \pos{43.9\%}
  & 84.3 & \pos{0.6\%} & {71.7} & \pos{1.4\%} & 79.3 & \pos{2.6\%}
  & {59.9} & \pos{4.4\%} & 58.4 & \pos{0.9\%} & {45.0} & \neg{-0.2\%} \\
\method- \tagSem
  & 88.3 & \pos{1.8\%} & {83.3} & \pos{6.4\%} & {\textbf{83.3}} & \textbf{\pos{56.3\%}}
  & {86.4} & \pos{3.1\%} & {\textbf{74.7}} & \textbf{\pos{5.7\%}} & 78.8 & \pos{1.9\%}
  & 58.9 & \pos{2.6\%} & {60.9} & \pos{5.2\%} & {45.0} & \neg{-0.2\%} \\
  
\rowcolor{gray!8}
\multicolumn{19}{l}{\textit{
Qwen3-Embedding-0.6B}} \\

\method- \tagStruc
  & 90.0 & \pos{3.8\%} & \textbf{88.3} & \pos{12.8\%} & 71.7 & \pos{34.5\%}
  & \textbf{87.4} & \textbf{\pos{4.3\%}} & 70.2 & \neg{-0.7\%} & \textbf{80.8} & \pos{4.5\%}
  & 59.9 & \pos{4.4\%} & 57.9 & \neu{0.0\%} & 46.5 & \pos{3.1\%} \\
\method-\tagRefl
  & 90.0 & \pos{3.8\%} & 81.7 & \pos{4.3\%} & 76.7 & \pos{43.9\%}
  & 84.8 & \pos{1.2\%} & 71.2 & \pos{0.7\%} & 77.8 & \pos{0.6\%}
  & 58.9 & \pos{2.6\%} & 57.9 & \neu{0.0\%} & 46.5 & \pos{3.1\%} \\
\method- \tagSem
  & 90.0 & \pos{3.8\%} & 81.7 & \pos{4.3\%} & 75.0 & \pos{40.7\%}
  & 85.4 & \pos{1.9\%} & \textbf{74.7} & \pos{5.7\%} & 78.8 & \pos{1.9\%}
  & 59.4 & \pos{3.5\%} & 59.4 & \pos{2.6\%} & 45.5 & \pos{0.9\%} \\
\bottomrule
\end{tabular}}
\caption{Impact of retriever choice. Comparison between Qwen3-Embedding-0.6B and e5-base-v2 on retrieval over \method-59k.}
\label{tab:qwen3emb}
\end{table*}

\subsection{Conflict Among Retrieved Traces}
\label{app:conflict}

Since we retrieve $k{=}3$ documents per query, it is possible that individually relevant traces conflict with one another once combined in context, producing a disjointed or contradictory set of retrieved guidance. To assess how often this occurs, we analyze the relationships among the top-3 traces retrieved for each AIME 2025--2026 query across all three \method{} variants. We use Gemini-2.5-Flash to classify each query's top-3 retrieved traces into one of three categories: \emph{Reinforcing} (the traces support a shared core approach), \emph{Complementary} (the traces offer distinct but compatible strategies), or \emph{Contradicting} (the traces recommend incompatible approaches).

\begin{table}[h!]
\centering
\footnotesize
\setlength{\tabcolsep}{2pt}
\renewcommand{\arraystretch}{1.0}
\begin{tabular}{@{}lccc@{}}
\toprule
\textbf{Variant} & \textbf{Reinforcing} & \textbf{Complementary} & \textbf{Contradicting} \\
\midrule
\tagStruc & 10.0\% & 90.0\% & 0.0\% \\
\tagSem   & 21.7\% & 78.3\% & 0.0\% \\
\tagRefl  & 31.7\% & 68.3\% & 0.0\% \\
\bottomrule
\end{tabular}
\caption{
Classification of top-3 retrieved traces across all 180 AIME 2025--2026 query--corpus pairs (60 queries $\times$ 3 \method{} variants), using Gemini-2.5-Flash as the classifier.
}
\label{tab:conflict-audit}
\end{table}

Across all 180 query--corpus pairs, no retrieved set is classified as contradicting (Table~\ref{tab:conflict-audit}). Retrieved traces are predominantly complementary, offering distinct but compatible strategies, or reinforcing a shared core approach. Notably, even \tagRefl, which is designed to expose alternative mistakes and corrective strategies and might therefore be expected to surface more disagreement, does not produce any genuinely conflicting retrieval sets. These results suggest that at $k{=}3$, retrieved traces tend to offer multiple compatible perspectives or repeated support for a common strategy rather than conflicting guidance. We note that this analysis uses an LLM-based classifier and is limited to AIME queries, so we treat it as an initial audit rather than a statistically conclusive estimate of conflict frequency across all tasks and retrieval settings.

\section{Decontamination Analysis}
\label{app:decontamination}
\begin{figure*}[t!]
    \centering
    \begin{subfigure}[t]{0.43\linewidth}
        \centering
        \includegraphics[width=\linewidth,valign=t]{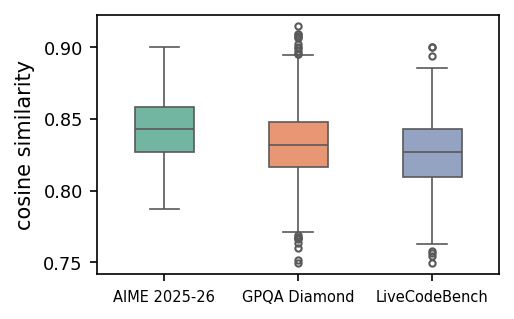}
        \label{fig:embed-sim}
    \end{subfigure}
    \hfill
    \begin{subfigure}[t]{0.49\linewidth}
        \centering
        \includegraphics[width=\linewidth,valign=t]{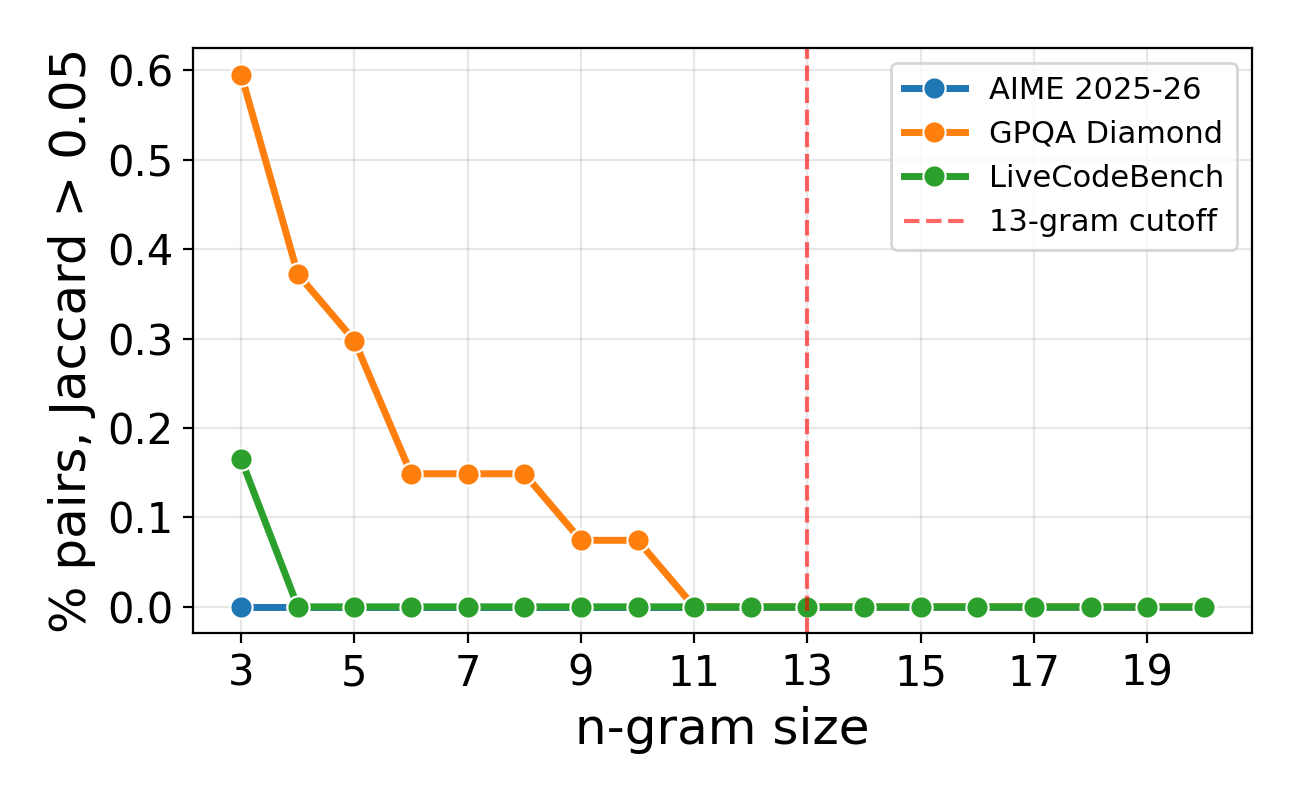}
        \label{fig:decontamination-ngram}
    \end{subfigure}
\caption{
Semantic and lexical contamination diagnostics on the raw  trajectory corpus. {Left:} \texttt{e5-base-v2} cosine similarity between benchmark queries and retrieved trajectory documents after 13-gram decontamination. {Right:} Percentage of retrieved query--passage pairs with $n$-gram Jaccard similarity above 0.05 as a function of $n$; the dashed line marks the 13-gram cutoff. High embedding similarity reflects semantic relatedness expected for retrieval, while long $n$-gram overlap drops to zero at the chosen threshold for all three benchmarks. Since this analysis uses raw trajectories, it is a conservative setting compared to the more compact transformed variants.}
    \label{fig:decontamination-analysis}
\end{figure*}
We perform additional checks to assess the risk of contamination between our trace corpora and evaluation benchmarks. Our first safeguard is temporal: some evaluation benchmarks, most notably AIME 2025--2026, were released after the source problem collections used to construct our trace corpora. This reduces the likelihood that exact benchmark problems appear in the auxiliary trace-generation sets. We nevertheless apply lexical decontamination following prior work \citep{borgeaud2022improving,lyu2025frustratinglysimpleretrievalimproves}, removing examples whose 13-gram Jaccard overlap with any evaluation query exceeds 0.05.

Figure~\ref{fig:decontamination-analysis} summarizes our lexical and semantic contamination checks on the raw Gemini-2-thinking trajectory corpus. We use raw trajectories for this analysis because they are the most surface-similar retrieval units; transformed traces are shorter and more abstract, and since they have been transformed, they form a less lexical contamination-prone setting. Figure~\ref{fig:decontamination-analysis} (left) shows the distribution of \texttt{e5-base-v2} cosine similarities between benchmark queries and retrieved trajectory documents. High embedding similarity is expected in a retrieval setting, since the retriever is designed to find conceptually related reasoning examples. We therefore treat embedding similarity as a diagnostic rather than direct evidence of contamination.

Figure~\ref{fig:decontamination-analysis}(right)  shows token $n$-gram Jaccard overlap between benchmark queries and retrieved trajectory documents, sweeping $n$ from 3 to 20. Lexical overlap drops rapidly as $n$ increases. At the 13-gram cutoff, no retrieved pairs exceed Jaccard $>0.05$ for AIME 2025--2026, GPQA-Diamond, or LiveCodeBench. This supports 13-gram Jaccard as a conservative near-duplicate filter: smaller $n$-grams can capture benign topical overlap or shared technical phrases, while longer $n$-grams are more indicative of copied surface form.

Overall, the analysis supports our decontamination procedure. The 13-gram filter removes long surface-form overlap, while the remaining high embedding similarity reflects the intended behavior of retrieval: finding semantically related but textually distinct reasoning traces. Since this analysis uses raw trajectories, it is conservative; transformed traces are more compact and abstract, making them less likely to preserve benchmark-specific surface form.

\begin{figure*}[t]
\centering
\small
\setlength{\tabcolsep}{8pt}
\renewcommand{\arraystretch}{1.3}
\begin{tabular}{@{}p{0.13\linewidth} p{0.73\linewidth} r@{}}
\toprule
& & \textbf{Tokens} \\
\midrule

\textit{Problem} &
$n$ coins are simultaneously flipped. The probability that at most one shows tails is $\frac{3}{16}$. Find $n$.
&  \\

\midrule

\textit{Full trace} &
The problem asks for the value of $n$, the number of coins flipped, given a probability condition on the number of tails observed. Let $X$ be the number of tails observed when $n$ coins are flipped. Each coin flip is an independent Bernoulli trial with probability of tails $p = 0.5$ \ldots\ldots\ldots \textit{[3,800 tokens omitted]} \ldots\ldots\ldots The reasoning used the properties of the binomial distribution and solving the resulting equation. The uniqueness of the solution was established by analyzing the monotonicity of the function involved.
& 3{,}867 \\

\midrule

\tagStruc &
\textbf{Approach:} Solving the Probability Equation\newline
\textbf{Step 1} --- Define $P(X \le 1) = (1+n)(0.5)^n$.\newline
\textbf{Step 2} --- Set $(1+n)(0.5)^n = \tfrac{3}{16}$; rewrite as $1 + n = 3 \cdot 2^{n-4}$.\newline
\textbf{Step 3} --- Test integer values; $n{=}5$ gives $1{+}5 = 3{\cdot}2^1 = 6$. \checkmark\newline
\textbf{Step 4} --- $(1+n)2^{-n}$ is strictly decreasing for $n \ge 1$, so the solution is unique.\newline
\textbf{Answer:} $\boxed{5}$
& 391 \\

\midrule

\tagSem &
\textbf{Pass 1 (full):} Let $X \sim \mathrm{Bin}(n,0.5)$. Then $P(X{\le}1)=(1+n)(0.5)^n=\tfrac{3}{16}$, giving $1+n=3\cdot 2^{n-4}$. Testing integers yields $n=5$; monotonicity of $(1+n)2^{-n}$ ensures uniqueness.\newline
\textbf{Pass 2 (core moves):} (i) Express $P(X{\le}1)$ via binomial PMF $\Rightarrow$ $(1+n)(0.5)^n = \tfrac{3}{16}$. \;(ii) Rewrite and solve by trial. \;(iii) Confirm uniqueness via monotonicity.\newline
\textbf{Pass 3 (insight):} Recognise the binomial condition; convert to $1+n=3\cdot 2^{n-4}$ and close by integer search.\newline
\textbf{Answer:} $\boxed{5}$
& 526 \\

\midrule

\tagRefl &
\textbf{Trap profile:} Binomial probability --- overlooking solution uniqueness.\newline
\textcolor{red}{$\times$}~\textbf{Assuming uniqueness} without proof: the problem says ``find $n$'' but a second solution could exist.\newline
\textcolor{red}{$\times$}~\textbf{Misapplying logarithms} to $(1+n)2^{-n}$, introducing or losing solutions.\newline
\textcolor{red}{$\times$}~\textbf{Forgetting the binomial coefficient} in $P(X{=}1) = n\cdot(0.5)^n$.\newline
\textcolor{teal}{$\checkmark$}~Substitute back into the \emph{original} equation, not a rearranged form.\newline
\textcolor{teal}{$\checkmark$}~Prove $(1+n)2^{-n}$ is strictly decreasing to establish uniqueness.\newline
\textbf{Correct approach:} Reduce to $1+n=3\cdot 2^{n-4}$, solve by trial, confirm uniqueness. \textbf{Answer:} $\boxed{5}$
& 630 \\

\bottomrule
\end{tabular}
\caption{A single reasoning trace transformed by each strategy, with token counts. All three transformations are substantially more compact than the full trace (3,867 tokens), while preserving the key reasoning content in different ways.}
\label{fig:transform-example}
\end{figure*}

\clearpage
\begin{figure*}[t]
\centering\small
\setlength{\tabcolsep}{8pt}
\renewcommand{\arraystretch}{1.3}
\begin{tabular}{@{}p{0.13\linewidth} p{0.73\linewidth} r@{}}
\toprule
& & \textbf{Tokens} \\
\midrule

\textit{Problem} &
A ball is launched with speed $v_0$ in a gravitational field. The boundary of the reachable region satisfies $z \le z_0 - kx^2$. Find $z_0$ and $k$.
& 139 \\

\midrule

\textit{Full trace} &
The problem asks for the constants $z_0$ and $k$ that define the boundary of the region reachable by a projectile launched with a fixed initial speed $v_0$. We are given the equation of the boundary \ldots\ldots\ldots \textit{[3,000 tokens omitted]} \ldots\ldots\ldots The condition when $z=0$ gives the maximum range on a flat surface, which also matches the known formula.
& 3{,}237 \\

\midrule

\tagStruc &
\textbf{Approach:} Envelope via Lagrange / angle optimisation\newline
\textbf{Step 1} --- Write $x=v_0\cos\theta\,t$, $z=v_0\sin\theta\,t - \tfrac{1}{2}gt^2$; eliminate $t$.\newline
\textbf{Step 2} --- Express $z = x\tan\theta - \tfrac{gx^2}{2v_0^2}(1+\tan^2\theta)$.\newline
\textbf{Step 3} --- Set $y=\tan\theta$; maximise $z$ over $y$: $\partial z/\partial y = x - \tfrac{gx^2}{v_0^2}y = 0 \Rightarrow y = v_0^2/(gx)$.\newline
\textbf{Step 4} --- Substitute back: $z_{\max} = \tfrac{v_0^2}{2g} - \tfrac{g}{2v_0^2}x^2$.\newline
\textbf{Step 5} --- Identify $z_0 = v_0^2/(2g)$, $k = g/(2v_0^2)$.\newline
\textbf{Answer:} $\boxed{z_0 = \tfrac{v_0^2}{2g},\; k = \tfrac{g}{2v_0^2}}$
& 494 \\

\midrule

\tagSem &
\textbf{Pass 1 (full):} Eliminate $t$ from parametric equations to get $z = x\tan\theta - \tfrac{gx^2}{2v_0^2}(1+\tan^2\theta)$. Set $y=\tan\theta$, differentiate with respect to $y$, and set to zero to get $y=v_0^2/(gx)$. Substituting back yields $z_{\max} = v_0^2/(2g) - gx^2/(2v_0^2)$, so $z_0=v_0^2/(2g)$ and $k=g/(2v_0^2)$.\newline
\textbf{Pass 2 (core moves):} (i) Derive trajectory $z(x,\theta)$ by eliminating $t$. (ii) Optimise over launch angle to find the envelope. (iii) Match coefficients to identify $z_0$ and $k$.\newline
\textbf{Pass 3 (insight):} The reachable boundary is the \emph{envelope} of all parabolic trajectories; maximising $z$ over the launch angle at fixed $x$ directly gives it.\newline
\textbf{Answer:} $\boxed{z_0 = \tfrac{v_0^2}{2g},\; k = \tfrac{g}{2v_0^2}}$
& 648 \\

\midrule

\tagRefl &
\textbf{Trap profile:} Projectile envelope --- forgetting to optimise over angle; vertex formula shortcut.\newline
\textcolor{red}{$\times$}~\textbf{Fixed-angle fallacy:} Applying kinematic range/height formulas for a single $\theta$ instead of finding the envelope over all angles.\newline
\textcolor{red}{$\times$}~\textbf{Vertex shortcut:} Using the parabola vertex for a fixed $\theta$ as the maximum height --- the vertex changes with $\theta$ and is not the boundary.\newline
\textcolor{red}{$\times$}~\textbf{Algebra error in elimination:} Dropping the $\sec^2\theta = 1+\tan^2\theta$ factor when substituting $t$, yielding a wrong trajectory.\newline
\textcolor{teal}{$\checkmark$}~Check $\theta=90^\circ$ (vertical launch): boundary gives $z=v_0^2/(2g)$, matching free-fall max height.\newline
\textcolor{teal}{$\checkmark$}~Check $z=0$: gives $x_{\max}=v_0^2/g$, consistent with the known range formula at $45^\circ$.\newline
\textbf{Correct approach:} Derive $z(x,\theta)$, treat $\tan\theta$ as a free parameter, maximise at fixed $x$, match to $z_0-kx^2$. \textbf{Answer:} $\boxed{z_0=\tfrac{v_0^2}{2g},\,k=\tfrac{g}{2v_0^2}}$
& 724 \\

\bottomrule
\end{tabular}
\caption{Example of generated and transformed thinking traces from a physics problem.}
\vspace{10em}
\label{fig:transform-physics}
\end{figure*}

\newpage
\begin{figure*}[t]
\centering\small
\setlength{\tabcolsep}{8pt}
\renewcommand{\arraystretch}{1.3}
\begin{tabular}{@{}p{0.13\linewidth} p{0.73\linewidth} r@{}}
\toprule
& & \textbf{Tokens} \\
\midrule

\textit{Problem} &
A text editor supports one operation: replace every occurrence of character $c_1$ with $c_2$. Given an input string and a target string, find the minimum number of operations to transform one into the other, or return $-1$ if impossible.
& 570 \\

\midrule

\textit{Full trace} &
Let's analyze the input and the problem statement carefully. For each position $i$ where $s_1[i] \neq s_2[i]$, we need a mapping $s_1[i] \to s_2[i]$ \ldots\ldots\ldots \textit{[3,900 tokens omitted]} \ldots\ldots\ldots Consider \texttt{ABCD} $\to$ \texttt{BACD}. A needs to become B, B needs to become A --- this requires 3 steps (an intermediate character is needed to break the cycle).
& 4{,}193 \\

\midrule

\tagStruc &
\textbf{Approach:} Transformation graph and connected components\newline
\textbf{Step 1} --- Build mapping $s_1[i] \to s_2[i]$; if any character maps to two different targets, return $-1$.\newline
\textbf{Step 2} --- Construct a directed graph: nodes are characters, edges are required transformations.\newline
\textbf{Step 3} --- Find connected components via DFS/BFS.\newline
\textbf{Step 4} --- Per component: if it contains a cycle, cost $= |\text{nodes}| + 1$ (one extra intermediate step); otherwise cost $= |\text{edges}|$.\newline
\textbf{Step 5} --- Sum costs across all components.\newline
\textbf{Answer:} minimum keystrokes, or $-1$ if impossible.
& 497 \\

\midrule

\tagSem &
\textbf{Pass 1 (full):} Build a character mapping from $s_1$ to $s_2$; any conflict (same source, two targets) is immediately $-1$. Represent mappings as a directed graph and find connected components. For each component, count cycles: each cycle of length $n$ needs $n+1$ operations (one intermediate); a simple path needs as many operations as edges.\newline
\textbf{Pass 2 (core moves):} (i) Detect contradictions while building the mapping. (ii) Model transformations as a directed graph. (iii) Count keystrokes as edges $+$ one extra per cycle.\newline
\textbf{Pass 3 (insight):} Cycles in the transformation graph are the key difficulty --- each requires an extra intermediate step to break; everything else is a simple chain.
& 389 \\

\midrule

\tagRefl &
\textbf{Trap profile:} String transformation graph --- mishandling cycles and global operation scope.\newline
\textcolor{red}{$\times$}~\textbf{Cycle cost off-by-one:} Counting a cycle of length $n$ as $n{-}1$ operations, missing the extra intermediate step needed to break it.\newline
\textcolor{red}{$\times$}~\textbf{Ignoring global scope:} Applying transformations position-by-position, forgetting that one keystroke affects \emph{every} occurrence of $c_1$.\newline
\textcolor{red}{$\times$}~\textbf{Missing contradictions:} Not checking whether one input character must map to two different targets at different positions.\newline
\textcolor{teal}{$\checkmark$}~Verify contradiction detection: does any source character appear with two different targets?\newline
\textcolor{teal}{$\checkmark$}~Trace a small cycle (e.g.\ A$\to$B$\to$A) manually to confirm the $+1$ intermediate step.\newline
\textbf{Correct approach:} Build transformation graph, detect conflicts, identify cycles, sum $|\text{edges}| + |\text{cycles}|$. Return $-1$ on contradiction.
& 620 \\

\bottomrule
\end{tabular}
\caption{Example of generated and transformed thinking traces from a coding / optimization example.}
\label{fig:transform-coding2}
\end{figure*}

\end{document}